\documentclass{article}              
\usepackage{arxiv}
\usepackage{amsmath}
\usepackage{bm}
\usepackage[dvips]{graphicx}
\usepackage{ulem}
\usepackage{epsfig}
\usepackage{float}
\usepackage{siunitx}

\usepackage{color}

\begin{document}                  



\title{Percus-Yevick structure factors made simple}


\author{Robert Botet \\
Laboratoire de Physique des Solides B\^{a}t.510, \\ CNRS UMR8502 / Universit\'{e} Paris-Saclay, Centre d'Orsay, F-91405 Orsay France \\
\And
Sylvie Kwok \\
CBI, ESPCI, CNRS, PSL University, 75005 Paris France \\
\And
Bernard Cabane \\
CBI, ESPCI, CNRS, PSL University, 75005 Paris France}









\maketitle                        

\begin{abstract}
Measuring the structure factor, $S(q)$, of a dispersion of particles by Small-Angle X-ray Scattering  provides a unique method to investigate the spatial arrangement of colloidal particles. However, it is impossible to find  the exact location of the particles from $S(q)$ because some information is inherently lacking in the SAXS signal. The two standard ways to analyse an experimental structure factor are then to compare it either to structure factors computed from simulated systems, or to analytical structure factors calculated from approximated systems. For liquids of monodisperse hard spheres, the latter method provides analytical structure factors through the Ornstein-Zernike equation used with the Percus-Yevick closure equation. The structure factors obtained in this way were not adequate for the more common dispersions of polydisperse particles. However, Vrij, Bloom and Stell were able to demonstrate that the same mathematical framework could be extended to yield accurate approximations for the experimental structure factor. Still, this solution has remained underused because of its mathematical complexity.
In the present work, we derive and report the complete Percus-Yevick solution for general polydisperse hard-spheres systems in a concise form that is straightforward to use. The form of the solution  is made simple enough to give ready solutions of several important particle-radius distributions (Schulz, truncated normal and inverse Gaussian). We also discuss in detail the case of the power-law radius distribution, relevant in the case of systems made of an Apollonian packing of spheres, as recently discovered experimentally in high internal phase ratio emulsions.  

\end{abstract}


\section{Introduction}

Small-Angle Scattering (SAXS for X-rays, SANS for neutrons) (Fejgin, 1987) is a powerful method to deduce structural information of materials (liquid or solid) in the range of distances $10$-$1000$ \si{\angstrom} (Guinier, 1939). This makes it a unique tool to study spatial organization of matter prepared for example, by dispersing fine particles (nano- or micrometer sized) in a liquid (Porod, 1951). 

In order to determine the particles' arrangement in such a material, we need to know the distance distribution functions for each type of pair of particles. This is generally a formidable problem, because the particles are not all identical, and also because intra-particle and inter-particle distances are confounded. As a result, the amount of information required to describe precisely the spatial structure far exceeds that available in the experimental data. There are several ways to deal with this problem:
\begin{itemize}
\item	change the contrasts of the particles without changing their positions, e.g. substituting H atoms with D atoms in SANS (Williams et al, 1979; Williams, 1991), or varying the wavelength of X-rays near the absorption edge of suitable atoms in SAXS (Stuhrmann, 1985). These contrast variation techniques can provide a rigorous separation of intra-particle and inter-particle distances, but are infrequently practised due to the experimental difficulty in ensuring structurally identical samples which scatter differently.
\item create a model that takes into account the distribution of particle sizes and shapes and use this model to predict the pair-distance distributions at equilibrium (Franke et al, 2017). This approach is feasible in cases where a sample is infinitely dilute, or if the particles are identical. However this is not generally the case because of the unknown interactions between particles of different sizes and shapes.
\end{itemize}

Half a century ago,  Baxter's solution to the Percus-Yevick equation in the case of an infinite number of hard-sphere species (Baxter, 1968; Baxter, 1970) suggested a way to solve the Ornstein-Zernike equation in this situation within the frame of the Percus-Yevick approximation (Percus \& Yevick, 1958). It was done for a system made of uncharged polydispersed hard spheres (Vrij, 1979; Blum\& Stell, 1979), and in particular for a distribution of diameters that matched the Schulz distribution (Vrij, 1978). This provides a complete mean-field solution for the structural problem at least for this particular case of a population of uncharged hard spheres (van Beurten \& Vrij, 1981; Griffith et al, 1986;Griffith et al, 1987). Several authors have since extended Vrij's solution to similar cases where a mean-field solution can be postulated (e.g. soft spheres (Blum \& Stell, 1979) or adhesive hard spheres (Robertus et al, 1989). All these solutions to the decomposition problem can be expected to be as good as the mean-field descriptions of phase transitions, which appear to be rather apt descriptors. Nevetheless, most users of small angle scattering have been deterred by the  complexity of the Vrij solution, and many have preferred to use an oversimplified solution such as assuming an infinitely dilute sample, an assumption that is not strictly correct in small-angle scattering.
Many other users have belittled the effects of polydispersity in systems that show fractionated crystallization, even though it is known that complex crystal structures are not incompatible with high polydispersity (Cabane et al, 2016).
															
We first show that the Vrij solution can be expressed in a form which is mathematically  simpler than the expressions known to date. We provide a few examples for the effects of intra- and interparticle interferences on the scattering patterns of moderately dilute dispersions, and compare these effects with those predicted according to other approximations. Finally we show that the intra- and interparticle correlations of very polydisperse hard-sphere systems can be predicted with great accuracy thanks to the use of Babinet's principle.

 From these examples, it becomes clear that the effects of polydispersity (size or interaction)  are extremely important in systems where packing tends to be a serious constraint, for instance, in highly concentrated emulsions. We believe that our simplified solution will greatly assist other researchers in deriving a much easier spatial interpretation of complicated real-life systems from their convoluted scattering data measured in the reciprocal space. 

\section{Simple form of $S(q)$ for polydisperse system in the PY-approximation} \label{Intro}

According to Vrij's work (Vrij, 1979; van Beurten \& Vrij, 1981), the structure factor $S(q)$ of a dispersion of
neutral spherical particles with distribution of radii $n(a)$ can be
calculated within the PY hard-sphere theory and expressed as a
combination of a finite set of functions averaged over the
distribution of radii. However, the
formalism through which these functions are expressed is quite
complicated, and this has strongly limited its usage.

In the present Section, we  propose  much simpler formulae in a concise form, and a few comments on the equations. The  details of the mathematical calculations  bridging the present results to previous works are found in  Appendix A.   \\

The system being polydisperse, one has first to introduce the averaging process, $\Big\langle \dots \Big\rangle$,  over the particle radius distribution $n(a)$. There are various possible definitions, and we consider in the present work the conventional notation (written respectively for a continuous and for a discrete radius distribution function):
\begin{align}
\Big\langle f(qa) \Big\rangle \equiv & \int_0^{\infty} f(qa) \, n(a) da ~, \label{def1} \\
\Big\langle f(qa) \Big\rangle\equiv & \sum_{\alpha} f(qa_{\alpha}) \, n(a_{\alpha}) ~. \label{def2}
\end{align}

Within the PY approximation, the structure factor of a dispersion of hard spheres with volume fraction $\phi$ is  written under the form (full details in Appendix A): 
\begin{align}
S(q) = \dfrac{Y/c}{X^2+Y^2} ~, \label{S1}
\end{align}
with the expressions for $X$ and $Y$:
\begin{align}
X = & \, 1+b + \dfrac{2 e f g+ d(f^2-g^2)}{d^2+e^2} ~, \label{X} \\
Y = & \,  c+ \dfrac{2d f g-e(f^2-g^2)}{d^2+e^2} ~, \label{Y} 
\end{align}
and the auxiliary quantities:
\begin{align}
\psi = & \, \dfrac{3\phi}{1-\phi} \label{psi} \\
b = &  \, \psi ~\dfrac{\Big\langle (\cos (q a)+q a \sin (q a))(\sin(q a)-q a \cos (q a)) \Big\rangle}{\Big\langle (q a)^3 \Big\rangle} ~, \label{b} \\
c = &  \, \psi ~\dfrac{\Big\langle (\sin(q a)-q a \cos (q a))^2 \Big\rangle}{\Big\langle (q a)^3 \Big\rangle} ~, \label{c} \\
d = & \, 1+ \psi ~\dfrac{\Big\langle (q a)^2 \sin (q a)\cos (q a) \Big\rangle}{\Big\langle (q a)^3 \Big\rangle} ~, \label{d} \\
e = & \,  \psi ~\dfrac{\Big\langle (q a)^2  \sin^2 (q a) \Big\rangle}{\Big\langle (q a)^3 \Big\rangle} ~, \label{e} \\
f = & \, \psi~ \dfrac{\Big\langle q a \sin(q a) (\sin(q a)-q a \cos (q a)) \Big\rangle}{\Big\langle (q a)^3 \Big\rangle} ~, \label{f} \\
g = &  -\psi ~\dfrac{\Big\langle q a \cos(q a)(\sin(q a)-q a \cos (q a)) \Big\rangle}{\Big\langle (q a)^3 \Big\rangle} ~. \label{g}
\end{align}

In the equations above, the radius distribution $n(a)$ does not need to be normalized since $S(q)$ finally depends on ratios of average functions of $qa$. The only mathematical condition allowing the solution's expression under the form of (\ref{S1})-(\ref{g}) is the finiteness of the moments $\Big\langle a^k \Big\rangle$ for $k = 0, 1, 2, 3$ (a condition that we henceforth suppose fulfilled).

An equivalent form -- of easier use in some cases -- involving functions of complex variables for the calculation of $S(q)$  is given in Appendix A, formulae (\ref{Sfin0})-(\ref{nu3}). \\

Practically, to calculate the value of the structure factor $S(q)$, one starts by calculating the seven averaged expressions required to compute the functions $b, c, d, e, f, g$ of $qa$, that is: $\Big\langle a^3 \Big\rangle$ and $\Big\langle a^k \sin 2qa \Big\rangle $, $\Big\langle a^k \cos 2qa \Big\rangle $ for $k = 0, 1, 2$. These averaged functions can be written in closed form in a number of cases. Then, one uses these functions in (\ref{X})-(\ref{Y}) to write the auxiliary functions $X$ and $Y$, and includes them at last in the expression (\ref{S1}) of the structure factor. 

An advantage of the expressions written above to calculate $S(q)$ is that all the averages are linear functions of the radius distribution. Therefore, if the studied distribution is (e.g.) bimodal and represented by the sum of two known unimodal distributions, $n(a) = n_1(a)+n_2(a)$, any average parameter  is the corresponding sums of the  averages for each distribution $n_1$ and $n_2$. \\

The expressions (\ref{S1})-(\ref{Y}) for the structure factor, and the relations giving the auxiliary functions $b,c,d,e,f$ and $g$, are valid for any system of non-overlapping spheres within the PY  hard-sphere model. Several comments about the formulae are noteworthy at this stage:
\begin{itemize}
\item
it is a simple exercise to rediscover the standard Percus-Yevick formula for the monodisperse case  (Kinning \& Thomas, 1984) after removing the average symbols $\Big\langle \cdots \Big\rangle$ in the formulae above. In this sense, the equations (\ref{S1})-(\ref{g}) are nothing but a complicated way to write the popular monodisperse hard-sphere Percus-Yevick solution in terms of trigonometric functions and powers of $qa$. 
\item
for any finite value of $\phi$, the auxiliary parameters for $q \rightarrow \infty$ are such that: $d \simeq 1$, $c,e,g \sim 1/q$ and $b, f$ behave as $\sim 1/q^2$, hence the known result: $\lim_{q \rightarrow \infty} S(q) = 1$. We can be more precise about that limit. Using the asymptotic expansions of $\int_0^{\infty} \cos( q a) n(a) da$ and of $\int_0^{\infty} \sin ( q a) n(a) da$ found by Olver (Olver, 1974), one finds from the above relations:
\begin{align}
S(q) \sim & \, 1+ \dfrac{K}{q^2} ~~~~~,~~\text{for}~q \rightarrow \infty ~, \label{asymptot} \\
K = &  \, \psi~ \dfrac{\Big\langle a \Big\rangle^3}{\Big\langle a^3 \Big\rangle} \left( 1+ \dfrac{\psi}{2} \dfrac{\Big\langle a^2 \Big\rangle^2}{\Big\langle a \Big\rangle \Big\langle a^3 \Big\rangle} \right)  ~. \label{K} 
\end{align}
In particular we deduce from (\ref{asymptot}) that the limit value $1$ is always approached from above (since $K >0$), with behaviour $\sim 1/q^2$ whatever the radius distribution. When the Percus-Yevick approximation is valid in a certain experimental situation, the relations (\ref{asymptot})-(\ref{K}) can then be used as the best fit of the corresponding experimental $S(q)$ profile at large $q$, and find at the same time an estimation of the combination (\ref{K}) of moments of the radius distribution. 
\item
for any radius distribution, one recovers the trivial result $\lim_{\phi \rightarrow 0} S(q) = 1$ (the case of no scattering). More precisely, the structure factor approximates at the first order in $\phi$, as:
\begin{align}
S(q) \simeq \dfrac{1}{1+2(b-f g/c)} ~~~~,~~ \text{for}~\phi \ll 1 ~. \label{approxSG}
\end{align}
An explicit example is discussed in the following Section \ref{expon}.
\end{itemize}

Before delving into the discussion of any particular case, let us mention here that the complete analytical forms of the auxiliary quantities (\ref{b})-(\ref{g}) are given explicitly below:
\begin{itemize}
\setlength\itemsep{0.em}
\item
 for the Schulz radius distribution: expressions after Equ.(\ref{Sc})
\item
for the truncated normal distribution: expressions after Equ.(\ref{no})
\item
for the inverse Gaussian distribution: expressions after Equ.(\ref{invg})
\item
for the power-law  distribution: expressions after Equ.(\ref{na})  
\end{itemize} 
 
Other radius distributions lead to analytical expressions of $b, c, d, e, f, g$. We leave it to the interested reader to calculate these expressions for other cases (e.g. the uniform distribution).

\section{A basic example: spheres with exponential radius-distribution} \label{expon}

To exemplify the formulae above, we discuss in this Section a simple, though non-trivial,  radius distribution of spheres, namely the  exponential distribution:
\begin{align} 
n(a) \propto e^{-a/\Big\langle a \Big\rangle} ~, \label{expo}
\end{align}
of average radius $\Big\langle a \Big\rangle$. It is a decreasing function of the radius $a$ over the entire range $0< a <\infty$, and its polydispersity ({\it i.e.} the ratio between standard deviation and average value) is: $p = 1$, that is quite a wide distribution. Its value for $a \rightarrow 0$ is finite, while its tail (for $a \rightarrow \infty$) is exponential. 

The exponential distribution (\ref{expo}) is  mathematically simple and is particularly relevant in a number of applications, e.g. colloidal aggregates in a viscous fluid (Bastea, 2006).

\subsection{The analytical PY solution for the exponential radius distribution} \label{exp1}

We use below the solution involving real-valued functions as written in the Section \ref{Intro}. Using the scaled variable $x \equiv 2q \Big\langle a \Big\rangle$, the six auxiliary functions $b, c, d, e, f, g$ are as follows:
\begin{align}
b = & \, \dfrac{\phi}{1-\phi} \dfrac{1+5x^2}{(1+x^2)^3} ~, \label{be} \\
c = & \, \dfrac{\phi}{1- \phi} \, x^3 \dfrac{5+x^2}{(1+x^2)^3} ~, \label{ce} \\
d = & \, 1+\dfrac{\phi}{1-\phi} \dfrac{3-x^2}{(1+x^2)^3} ~,  \label{de} \\
e = & \, \dfrac{\phi}{1-\phi} \, x \dfrac{6+3x^2+x^4}{(1+x^2)^3} ~, \label{ee} \\
f = & \, \dfrac{\phi}{1-\phi} \, x^2 \dfrac{5+x^2}{(1+x^2)^3} ~, \label{fe} \\
g = & \, \dfrac{\phi}{1-\phi} \, x \dfrac{-2+3x^2+x^4}{(1+x^2)^3} ~, \label{ge} 
\end{align}
and they are all ratio of simple polynomials in $x$. The relations (\ref{X})-(\ref{Y}) and consequently (\ref{S1}) being rational functions of $b, c, d, e, f, g$, we obtain a simple formula for $S(q)$, namely:
\begin{align}
S(q) = & (1- \phi)^2 \dfrac{(1+x^2)^3}{5+x^2} \dfrac{P_2(x^2)}{Q_4(x^2)} ~, \label{Sexp} 
\end{align}
in which $P_2$ is a quadratic polynomial of its argument, $Z \equiv x^2$, and $Q_4$ a quartic polynomial, both with coefficients depending only on the value of $\phi \,$ namely:
\begin{align}
P_2(Z) = & \, 5+4\phi^2+2(3-5\phi+3 \phi^2) Z+(1-\phi)^2 Z^2 ~, \nonumber \\
Q_4(Z) = & \, (1+2\phi)^2+4(1+5 \phi^2) Z+ \nonumber \\
+ & 2(3-8\phi+9\phi^2-2 \phi^3) Z^2+ \nonumber \\
+& 4(1-\phi)^4 Z^3+(1-\phi)^4 Z^4 ~. \label{Q4}
\end{align}
All the coefficients of $P_2$ and $Q_4$ are positive for any value of the volume fraction, $0 \le \phi \le 1$. The asymptotic behaviour of $S(q)$ given by (\ref{Sexp})-(\ref{Q4}) for the large values of $x$ is consistent with the formulae (\ref{asymptot})-(\ref{K}), that is:
\begin{align}
(q \Big\langle a \Big\rangle)^2 \left( S(q) -1 \right) \simeq \dfrac{\phi}{2(1- \phi)^2} ~~~~, \text{for}~~~~ x \rightarrow \infty \label{asexp}
\end{align}
with the constant in the right-hand side of (\ref{asexp}) related to moments of the radius distribution through the expression (\ref{K}).\\

Although we are mainly interested in the present work in dense polydisperse systems, we note here the particular case of small values of $\phi$ for which the PY approximation is expected to be correct. One finds from (\ref{Sexp}), or using the approximation (\ref{approxSG}):
\begin{align}
S(q) \simeq \dfrac{1}{1 +2\phi\dfrac{3-x^2}{(1+x^2)^2}} ~~,~~ \text{when}~\phi \ll 1 ~, \label{approxS}
\end{align} 
which ensures that $S(q) \equiv 1$ for $\phi = 0$, as it must be. The difference between the approximation (\ref{approxS}) and the formula (\ref{Sexp}) is less than $5\%$ when $\phi < 0.1$.
An interesting behaviour of  (\ref{approxS}) is that, for this exponential radius distribution, the peak of $S(q)$ is asymptotically located at $q \Big\langle a \Big\rangle = \sqrt{7}/2 \simeq 1.32$ for $\phi \rightarrow 0$, while  the corresponding peak height behaves as $S_{\text{max}} = 1+ \phi/8$. 

This behaviour and these values can be compared to what happens to a system of monodisperse hard spheres of common radius $\Big\langle a \Big\rangle$, at  small volume fraction $\phi$, in the PY approximation. Indeed, in the monodisperse case, the location of the first peak of $S(q)$  is such that: $q \Big\langle a \Big\rangle \simeq 2.882$ (that is the smallest positive root of the equation $ \tan (2x) = 6x/(3-4x^2)$), when $\phi \rightarrow 0$, while the corresponding peak height is: $S_{\text{max}} \simeq 1+ \phi/1.45$, that is much larger than the polydisperse case. Incidentally, this is a sign that spatial correlations are stronger in the monodisperse than in the polydisperse case, even in the low-$\phi$ domain.

\subsection{Comparison of the PY solution with numerical simulations of systems of hard spheres with exponential radius distribution}

In this Section, we compare the PY polydisperse structure factor with structure factors obtained from simulated systems in order to specify the range of practical use of the approximation. Then, we introduce the Complementary Percus-Yevick hard-spheres approach to calculate approximately the structure factor of dense systems through Babinet's principle. 

\subsubsection{Comparison with numerical simulations of random dispersions of spheres} \label{simul}

The PY approximation supposes the same local radius distribution around each particle regardless of its size (Greene et al, 2016). This is a mean-field approximation.  Sticking to this constraint, we used Monte-Carlo simulation in the semi-grand canonical ensemble (Briano \& Glandt, 1984), putting randomly, in a cubic box with periodic boundary conditions, $N$ non-overlapping spheres selected with  radii following the exponential distribution (\ref{expo}). That way, we enforce  particle size and position to be essentially uncorrelated (except volume exclusion), since the next particle to add is sensitive to the remaining void geometries and not directly to the  particles surrounding the voids. Using these rules, we generated a number of independent systems comprising $N = 25\,000$ exponentially distributed spheres at controlled  volume fraction $\phi$. Using this algorithm, the maximum attainable volume fraction is $\simeq 0.69$.

\begin{figure}
\centering
\fbox{
\includegraphics[scale=0.20]{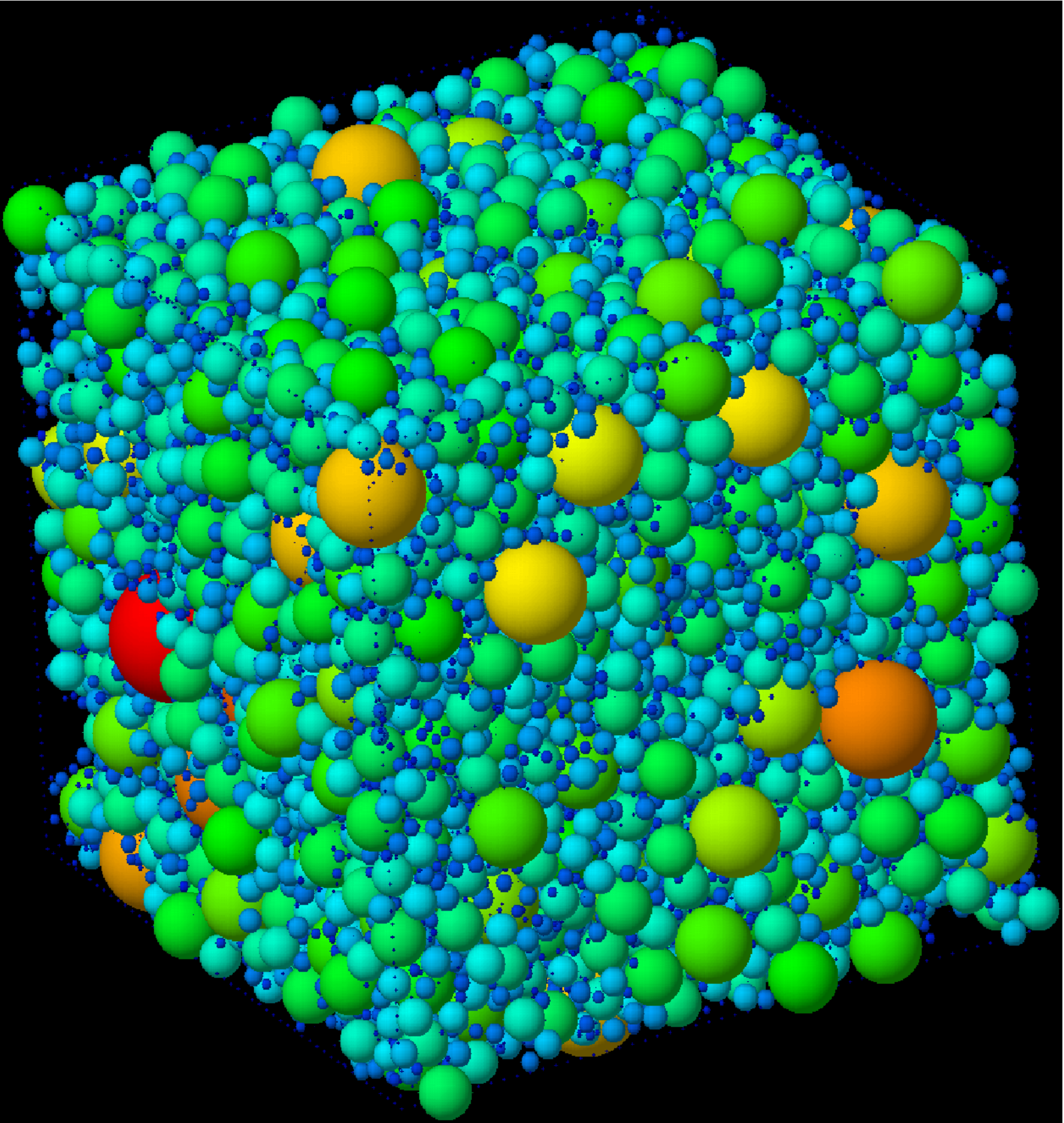}
\includegraphics[scale=0.25]{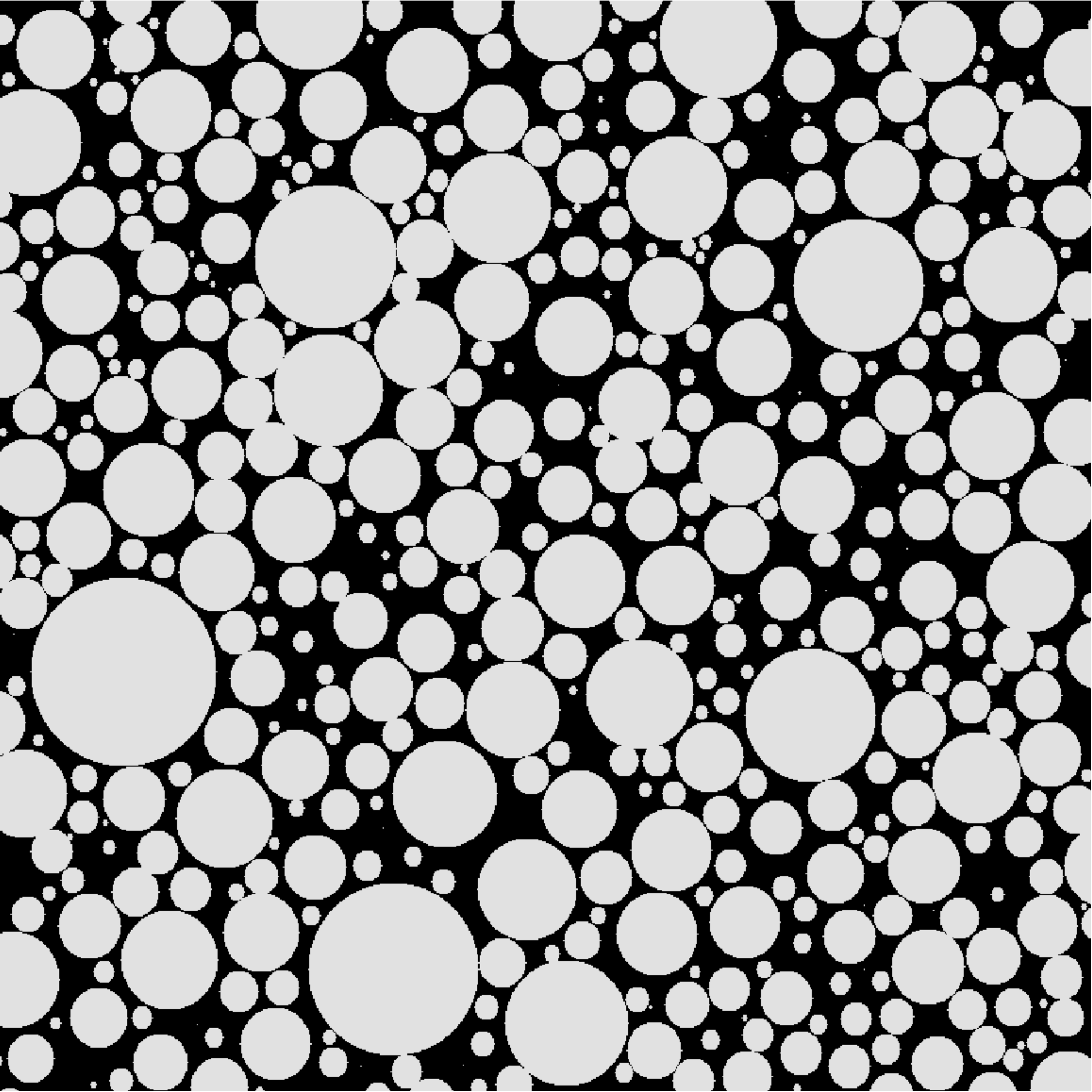}}
\caption{(left, colours on line): sketch of a 3D system of $N = 25\,000$ spheres with radii exponentially distributed, with average radius $\Big\langle a \Big\rangle$, packed at volume fraction $\phi = 0.69 $ in a cubic box with periodic boundary conditions. This is approximately the maximum volume fraction possible with the algorithm used (see text). Colours represent spheres of different radii ; (right): a cross-section of thickness $\Big\langle a \Big\rangle$ through the 3D system. }
\label{figsection}
\end{figure}

An example of such a system is presented in FIG.\ref{figsection} \\

\begin{figure}
\vspace*{-\baselineskip}
\centering
\includegraphics[scale=0.35]{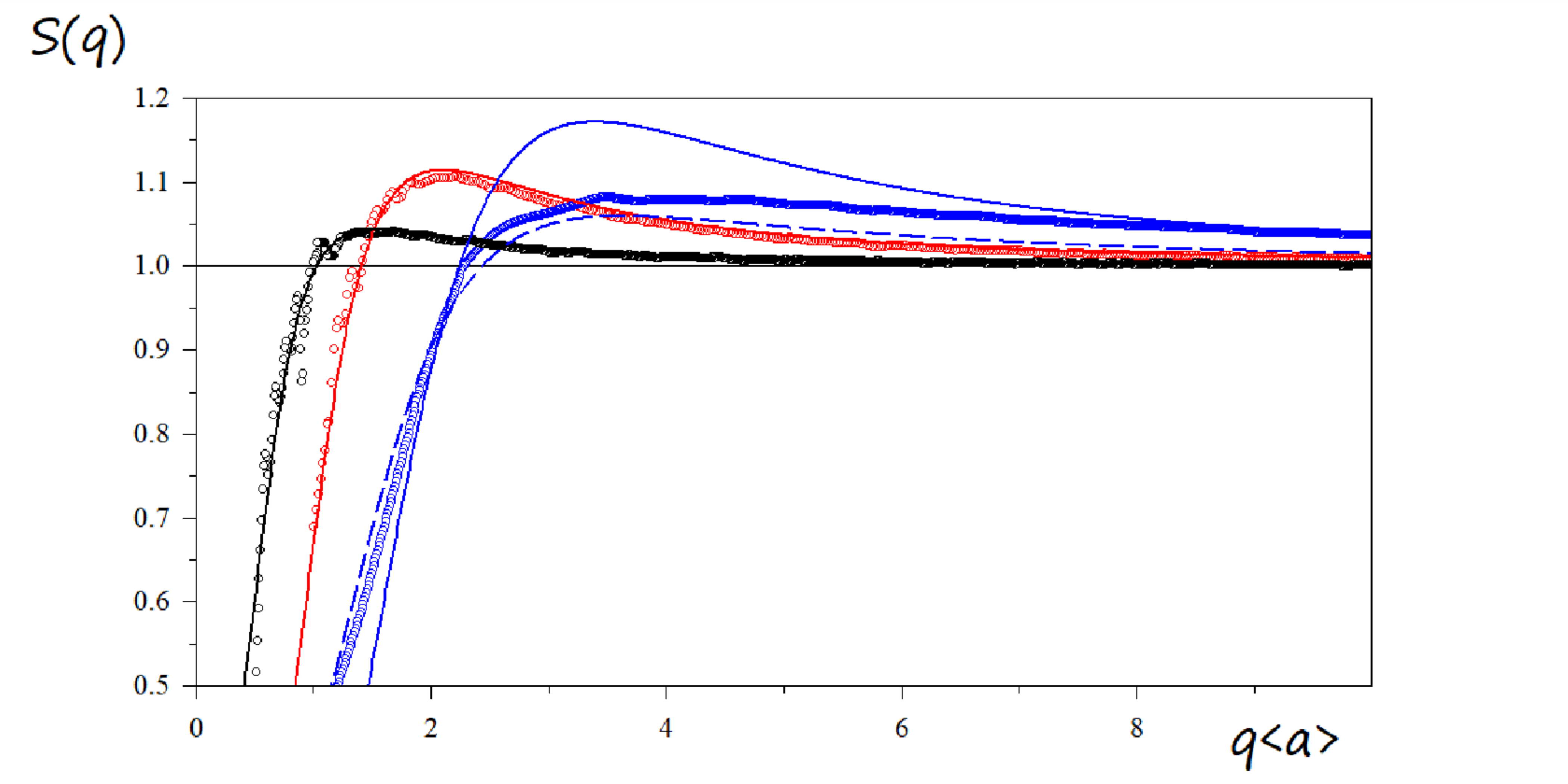}
\caption{Structure factors (circles)  of simulated systems constituting $N = 25\,000$ non-overlapping spheres in a cubic box with periodic boundary conditions. The radii of the spheres are distributed according to the exponential function (\ref{expo}), and the volume fractions are (from left to right): $\phi = 0.25 $ (black circles), $\phi = 0.50$ (red circles) and $\phi = 0.69 $ (blue circles). The results here shown are the average values taken over  11 independent samples for each volume fraction. The numerical data  are compared with the respective PY solutions (the formula (\ref{Sexp})) for the exponential radius distribution and same volume fractions (the continuous curves, respectively black, red and blue). Agreement between the PY solution and the numerical data is excellent up to $\phi = 0.50$. For $\phi > 0.50$, the numerical peak is significantly flattened compared to the analytical data. The dashed blue curve is the analytical structure factor based on Babinet's principle (see text) of a population of  voids at volume fraction $1-\phi = 0.31$, with void sizes exponentially distributed.}
\label{figexpon}
\end{figure}

In FIG. \ref{figexpon}, we show the structure factors, both numerical (red circles) and analytical (black continuous curves), in the cases: $\phi = 0.25$, $\phi = 0.50$ and $\phi = 0.69$.  

Agreement between the structure factor functions from the simulated systems and the corresponding analytical solutions from the PY approach is almost perfect up to $\phi = 0.5$. Beyond this value, the numerical peak height decreases with the volume fraction, while the analytical peak continues to grow. This discrepancy  beyond $\phi \simeq 0.5$ may be interpreted as specific spatial correlations and sizes correlations which are presumed in the PY approximation, whereas these presumptions are not present in the random-addition numerical simulations. 

\subsubsection{the Complementary Percus-Yevick hard-sphere approach for dense sphere packings} \label{CPY}

The discrepancy above can be roughly explained by the swapping of roles between matter and voids: when $\phi > 0.50$, the dense system of packed spheres may be considered as a population of voids inside a homogeneous medium. A pore is here defined as an interstice between neighbouring spheres. Each pore is an assembly of spherical triangles,   possibly connected to other pores by necks not wider than a threshold (e.g. the minimal sphere radius). Because of Babinet's principle, the definite structure factor is reduced to the X-ray scattering by an ensemble of small void domains in homogeneous matter (Guinier et al, 1955). 

To be more precise, although the shapes of the voids are  generally complicated, we  know exactly the average size of the voids. Indeed, if $v$ and $s$ denote respectively the volume and the surface of a given void, the following general ({\it i.e.} valid for any radius distribution) equations relate the total surface, $S$, of the interface between void and matter, and the volume fraction, $\phi$, of matter:
\begin{align}
1- \phi = \frac{N_v}{V} \Big\langle v \Big\rangle_{v} ~~~~;~~~~  \phi = \frac{N_s}{V} \frac{4\pi}{3} \Big\langle a^3 \Big\rangle ~, \label{fifi} \\
\frac{S}{V} = \frac{N_v}{V} \Big\langle s \Big\rangle_{v} ~~~~;~~~~  \frac{S}{V} = \frac{N_s}{V} 4 \pi \Big\langle a^2 \Big\rangle ~, \label{ss}
\end{align}
in which $N_v$ (resp. $N_s$) is the total number of voids (resp. spheres) in the volume $V$, and $\Big\langle \cdots \Big\rangle_{v}$ denotes the average value for the ensemble of the voids.  One can then eliminate $N_V, N_s$ and the volume $V$ from the equations above to deduce the relation between the typical sizes of the voids and the typical sizes of the spheres:
\begin{align}
\frac{\Big\langle v \Big\rangle_{v}}{\Big\langle s \Big\rangle_{v} } = \frac{1}{\psi} \frac{\Big\langle a^3 \Big\rangle}{\Big\langle a^2 \Big\rangle } ~. \label{voidsz}
\end{align} 
Let us define the effective radius, $a_{\text{eff}}$, of the void of volume $v$, by the relation: $v = 4 \pi a_{\text{eff}}^3/3$ (this means that the form factor of the void is replaced by the form factor of a sphere of same volume). The probability distribution of the $a_{\text{eff}}$ is not known, but the voids are essentially uncorrelated  (Brownlee, 1965), then we can invoke a theorem stating that the maximum entropy probability distribution for a positive random variate with fixed average value is the exponential distribution (Park \& Bera, 2009). Then, we assume that the void effective radii are distributed exponentially, with:
\begin{align}
\Big\langle a_{\text{eff}} \Big\rangle_{v} =  \frac{1- \phi}{\phi} \frac{\Big\langle a^3 \Big\rangle}{\Big\langle a^2 \Big\rangle } ~. \label{aeff}
\end{align}  
The last step is to suppose that the Percus-Yevick theory is valid for the ensemble of the voids, that is to say: the voids behave as an ideal gas with a pair distribution function which vanishes for distance $< a_{\text{eff}}$.   
Then, we can use the formula (\ref{Sexp}) to obtain the structure factor of the system, in which one replaces the matter volume fraction $\phi$ by the complementary void volume fraction $1 - \phi$. We call this method the Complementary Percus-Yevick hard-sphere approach.

This approximation results in the structure factor shown as the dashed blue curve in  FIG. \ref{figexpon} for the case of $\phi = 0.69 $. This structure factor, obtained from \mbox{X-ray} scattering by a dilute population of voids with an exponential distribution of their effective radius, is  much closer to the structure factor obtained from the simulated systems than the Percus-Yevick solution for scattering by $\phi = 0.69 $ of matter. \\

\begin{figure}
\vspace*{-\baselineskip}
\centering
\includegraphics[scale=0.9]{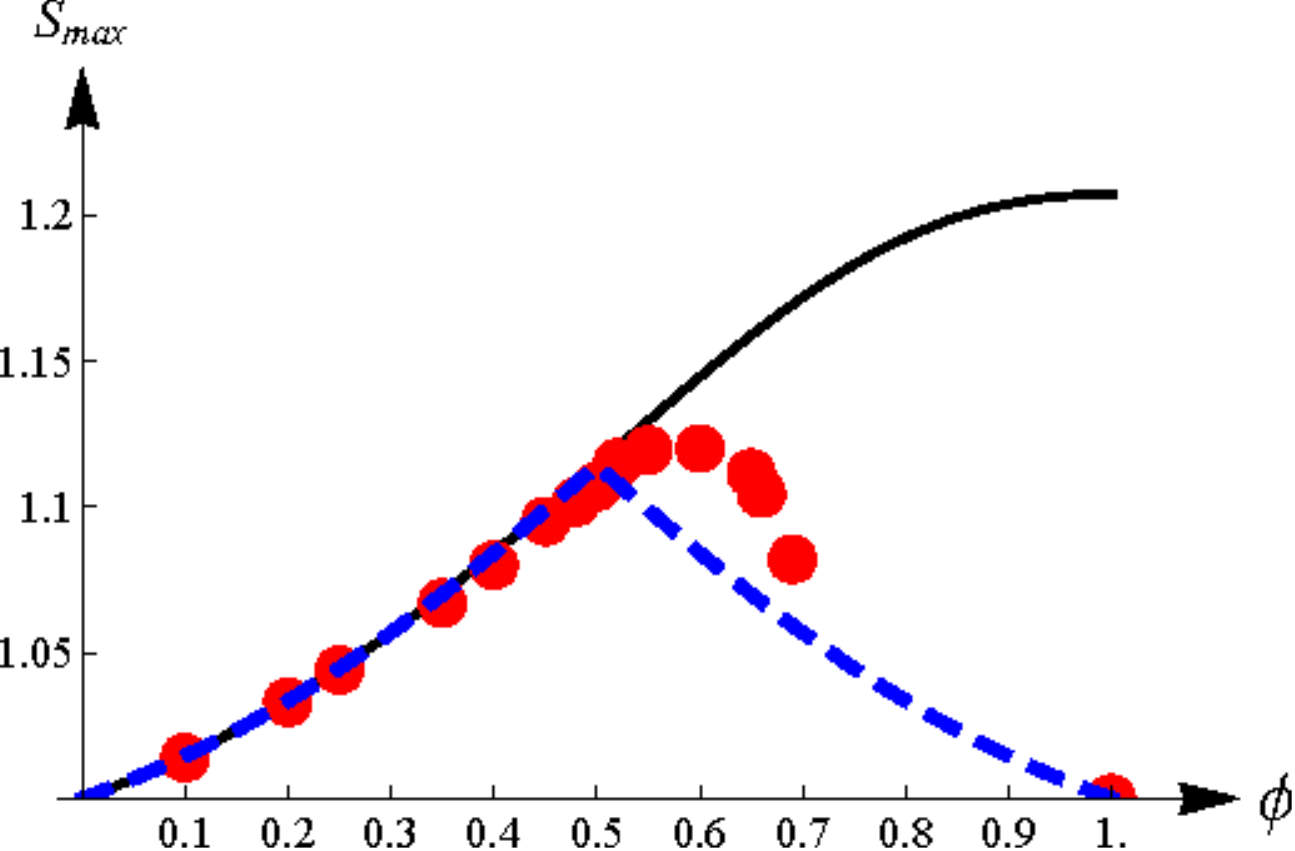}
\caption{Within the PY hard-sphere approximation, the height, $S_{\text{max}}$,  of the structure factor peak for the exponential radius distribution (\ref{expo}), is an ever-increasing function of the volume fraction $\phi$ (black continuous curve). This  analytical PY solution is here compared to the values of $S_{\text{max}}$ from simulated systems (red dots; see caption of Fig.\ref{figexpon} for details on the numerical simulations): up to $\phi \simeq 0.50$, numerical and analytical data agree perfectly well, also clearly illustrated in FIG. \ref{figexpon}. When $\phi > 0.50$, $S_{\text{max}}$ of the simulated systems are lower than one would expect from the PY solution at the same volume fraction. This non-monotonic behaviour of $S_{\text{max}}$ of the simulated systems is instead well reproduced using the Complementary Percus-Yevick solution (dashed blue curve), which only underestimates $S_{\text{max}}$ by no more than  $3 \%$. Note: the packing algorithm used to generate the simulated systems functioned only up to  $\phi > 0.69$ (see Section \ref{simul}). The point $(\phi = 1, S_{\text{max}} = 1)$ represents the case of scattering by pure homogeneous matter. }
\label{figsx}
\end{figure}

The complementary PY approach is expected to be general for dense dispersions of uncharged hard spheres, leading to  accurate estimations of the analytical structure factors when  volume fraction is so large that  voids are well represented as small isolated and uncorrelated scattering domains  randomly dispersed in homogeneous matter. 
Using such an approach, the agreement between the simulated and analytical data is much better (quantitatively better than $3\%$), as seen in FIG. \ref{figsx}.

\section{Exact analytical $S(q)$ for some radius distributions within the PY~approximation} \label{analy}

$S(q)$'s calculation is essentially reduced to determining the averaged auxiliary functions $ b, c, d, e, f, g$ defined after (\ref{S1})-(\ref{Y}). Then, the expression of the structure factor in the PY approximation is analytical as soon as the averages $\Big\langle a^k e^{i q a} \Big\rangle$ of the radius distribution $n(a)$ are known in closed forms for $k = 0, 1, 2, 3$. Hereafter is a limited list of  standard (and useful) 2-parameters continuous distributions for which the structure factor is calculated exactly. These full formulae can be used, for example, to fit  shapes of experimental structure factors resulting from SAXS analysis of polydisperse sphere dispersions. 

\begin{itemize}
\item
{\bf the Schulz distribution}
\begin{align}
\boxed{n(a) \propto  ~ a^{s-1} e^{-s \, a/\langle a \rangle}} ~,  \label{Sc} 
\end{align}
with $s>0$, $\Big\langle a \Big\rangle > 0$, and the radius, $a$, takes real positive values. The $S(q)$ for this distribution was already given in (van Beurten \& Vrij, 1981; Griffith et al, 1986; Griffith et al, 1987), and we present a much simpler form below.

The parameter $s$ is related to the polydispersity of the distribution, $p$,  through:
\begin{align}
p = \frac{1}{\sqrt{s}} ~. 
\end{align}
The behaviour of the distribution near $a \rightarrow 0$ is a power law, either decreasing ($\sim 1/a^{1-s}$) when $s <1$, or increasing ($\sim a^{s-1}$) when $s >1$. The Schulz distribution belongs to the exponential family, {\it i.e.} the tail of the distribution follows $\sim \exp(-a)$. 
The pure exponential distribution detailed in the previous section represents the case where $s = 1$. 
 
The structure factor $S(q)$ can be written explicitly in terms of functions of the reduced variable:
\begin{align}
x \equiv \frac{2q \Big\langle a \Big\rangle}{s} ~. \label{rvs}
\end{align}
Some combinations of variables appear naturally in the final expressions of $S(q)$ in this case. These are  the two trigonometric quantities:
\begin{align}
\Gamma \equiv & \, \dfrac{ \cos((s+2 ) \arctan(x))}{(1 + x^2)^{1 + s/2}} ~, \label{Delta} \\
\Sigma \equiv & \, \dfrac{ \sin((s+2 ) \arctan(x))}{(1 + x^2)^{1 + s/2}} ~, \label{Sigma} 
\end{align}
and three simpler quantities:
\begin{align}
g_1 \equiv & \,  \dfrac{\psi}{(s + 2) x} ~, \label{g1} \\
g_2 \equiv & \, \dfrac{2 \psi}{(s + 1) (s + 2) x^2} ~, \label{g2} \\
g_3 \equiv & \, \dfrac{4 \psi}{s (s + 1) (s + 2) x^3} ~, \label{g3}
\end{align} 
with the parameter $\psi$ given by (\ref{psi}).
Using $\Gamma, \Sigma, g_1, g_2, g_3$, the auxiliary quantities $b, c, d, e, f$ and $g$ are written exactly as:
\begin{align}
\boxed{
\begin{array}{ll}
b = & \, \left(g_3 - \dfrac{s + 4}{s} g_1 \right) \Sigma - 2 \dfrac{s + 2}{s} g_2 \Gamma  ~, \nonumber \\
~ &  ~\nonumber \\
c = & \,  g_1 + g_3 - \left( g_3 - \dfrac{s + 4}{s} g_1 \right) \Gamma - 2 \dfrac{s + 2}{s} g_2 \Sigma ~,  \nonumber \\
~& ~ \nonumber \\
d = & \, 1 +  g_1 \Sigma ~, \nonumber \\
~&  ~\nonumber \\
e = & \,  g_1 - g_1 \Gamma ~,  \nonumber \\
~&  ~\nonumber \\
f = & \, g_2  - g_2 \Gamma - \dfrac{s + 3}{s+1} g_1 \Sigma ~, \nonumber \\
~&  ~\nonumber \\
g = & \,   g_1 - g_2 \Sigma + \dfrac{s + 3}{s+1} g_1  \Gamma ~.  \nonumber
\end{array}}
\end{align}
The corresponding structure factor $S(q)$ for the Schulz distribution is calculated using  (\ref{X})-(\ref{Y}) then (\ref{S1}). 

{\it Note}: when $s$ is an integer number, $\Gamma$ and $\Sigma$ are ratios of polynomials in $x$. Because of (\ref{S1})-(\ref{Y}), the structure factor expresses in this case as the ratio of two polynomials of degree $2+3s$ in $x^2$. 

\item
{\bf the truncated normal distribution}
\begin{align}
\boxed{n(a) \propto \, e^{-\, \cfrac{(a-\langle a \rangle)^2}{2 p^2 \langle a \rangle^2}}} ~,  \label{no}
\end{align}
with $\Big\langle a \Big\rangle > 0$, and $p$ is the polydispersity of the distribution. 
The values of the radius $a$ must be restricted to the positive axis (hence the ``truncation'' of the full distribution). 

The value of the distribution near $a = 0$ is finite positive, while the tail  is Gaussian: $\sim \exp (-a^2)$. 

Analytical calculation of the auxiliary quantities $b, c, d, e, f, g$, requires consideration of  unphysical negative values of  particle radii $a$.  
The relative amount of these radii is the complementary error function $\text{Erfc}(1/p\sqrt{2})/2$ that we have to add to the truncated probability distribution. A rough estimation of the accuracy of the expressions below is given by the argument that for $p < 43\%$, Erfc( ) amounts to less than $1\%$. This will be also the  precision for $S(q)$ calculated by the current approach, provided the polydispersity $p$ is restricted to: $0 < p < 0.43$.  Within these conditions, using the scaled variable: 
\begin{align}
x \equiv 2 q \Big\langle a \Big\rangle ~, 
\end{align}
the auxiliary functions $\Gamma$ and $\Sigma$ defined as:
\begin{align}
\Gamma \equiv & \dfrac{3\phi}{(1+3p^2)(1-\phi)x^3} e^{-p^2 x^2/2} \cos x ~, \nonumber \\
\Sigma \equiv & \dfrac{3\phi}{(1+3p^2)(1-\phi)x^3} e^{-p^2 x^2/2} \sin x ~, \nonumber 
\end{align}
and the three simpler quantities:
\begin{align}
g_1 \equiv  & \, \dfrac{(1 + p^2)\psi }{(1+3p^2)x} ~, \nonumber \\
g_2 \equiv & \, \dfrac{2 \psi}{(1+3p^2)x^2} ~, \nonumber \\
g_3 \equiv & \, \dfrac{ 4 \psi}{(1+3p^2)x^3} ~. \nonumber
\end{align}
Using $\Gamma, \Sigma, g_1, g_2, g_3$, the auxiliary quantities $b, c, d, e, f$ and $g$ write:
\begin{align}
\boxed{
\begin{array}{ll}
b = & \,  ((2 + p^2 x^2)^2 -  x^2 (1 + p^2)) \Sigma - 2 x (2 + p^2 x^2) \Gamma ~, \nonumber \\
~& \nonumber \\
c = & \,  g_1+g_3 - 2 x (2 + p^2 x^2) \Sigma - ((2 + p^2 x^2)^2 - x^2 (1 + p^2)) \Gamma ~, \nonumber \\
~& \nonumber \\
d = & \, 1 + x^2  (1 + p^2 - p^4 x^2) \Sigma + 2 p^2 x^3 \Gamma ~, \nonumber \\
~& \nonumber \\
e = & \, g_1 +  2 p^2 x^3 \Sigma - (1 + p^2 - p^4 x^2) x^2\Gamma ~,  \nonumber \\
~& \nonumber \\
f = & \, g_2 - 2 x(1 + p^2 x^2) \Gamma - x^2 (1 - p^2 - p^4 x^2) \Sigma ~,  \nonumber \\
~& \nonumber \\
g = & \, g_1 + x^2 (1 - p^2 - p^4 x^2) \Gamma -  2x (1 + p^2 x^2) \Sigma ~. \nonumber
\end{array}} \nonumber
\end{align}
As usual, the corresponding structure factor $S(q)$ for the truncated normal distribution is calculated using  (\ref{S1})-(\ref{Y}). \\

In the case where the polydispersity $p > 43\%$ (that is the case where integration over positive radius should not be replaced by integration over the whole real axis), the respective formulas are more complicated, though still analytical, provided that the auxiliary quantities $\Gamma$ and $\Sigma$ are expressed in terms of the two complementary error functions $\text{Erfc}[(1/p  \pm  i p \, x)/\sqrt{2}]$.  The full expressions 	are not given here, but they are easy to obtain. 

\item 
{\bf the inverse Gaussian distribution}:
\begin{align}
\boxed{n(a) \propto \frac{1}{a^{3/2}} \, e^{-\cfrac{ (a -\langle a \rangle)^2}{2 p^2 \langle a \rangle \, a}}} ~,  \label{invg}
\end{align}
with $\Big\langle a \Big\rangle > 0$, and $p$ is the polydispersity of the distribution. 
The radius $a$ takes real positive values. 

The distribution belongs to the exponential family, and it  goes  sharply to 0 ($\sim \exp(-1/a)$) for $a \rightarrow  0$. In the context of the colloids, it is not as well-known as the Schulz and the normal distributions, though one can foresee generic conditions where the inverse Gaussian distribution may appear. Indeed, from the works of Schr\"odinger and Smoluchowski (Schr\"odinger, 1915; von Smoluchowski, 1915), we know that this distribution characterizes the time for a particle undergoing Brownian motion with a drift, to cover a given distance along a line. One has then to think of  a population of such particles growing at a  constant rate in a limited space domain, to obtain an inverse Gaussian size distribution. For example, it could be the case of micrometric particles sedimenting in a dispersion of tiny nanometric particles and accreting them (Alexandrov \& Lacis, 2000). In a different context, the inverse Gaussian distribution has been used to describe  the cell size distribution of phytoplankton population (Bernard et al, 2007). 

The exact solution, in this case, is much simpler when expressed in terms of functions of complex-valued variables, rather than using the real-valued quantities $b, c, d, e, f, g$. We use the formulae written in  (\ref{mu1})-(\ref{nu3}) (in Appendix 1) to calculate the  expressions for $\mu_1, \mu_2, \mu_3$ defined below. 

Firstly, we introduce the auxiliary quantity $\mu_0$:
\begin{align}
\mu_0 = & \,\dfrac{1}{1-4 i x p^2}\,  e^{(1-\sqrt{1-4 i x p^2})/p^2} ~, \nonumber
\end{align}
in which  the reduced variable, $x$, is:
\begin{align}
x \equiv q \Big\langle a \Big\rangle ~. \label{xig}
\end{align}
One then  obtains:
\begin{align}
\boxed{
\begin{array}{rcl}
\mu_{1} =  \, \dfrac{i}{2} \Big[ 1 +  \left(1+p^2 \right) x^2 -  ~~~~~~~~~~~~~~~~~~~~~~~~~~~~~~\\ 
-  \mu_0 \left(1 - 4 i x p^2 -  x^2 - i x \dfrac{  2  - 9 i x p^2}{\sqrt{
  1 - 4 i x p^2}}  \right) \Big] ~, \\[4mm] 
\mu_{2} =  \, \dfrac{i x^2}{2} \left( 1 + p^2 - \mu_0 \left(1 + \dfrac{p^2}{\sqrt{1 - 4 i  x p^2}} \right) \right) ~, ~~~~~ \\[4mm] 
\mu_{3}=  \, \dfrac{x}{2} \Big( 1 + i x (1+p^2)  + \mu_0 (i x -  \dfrac{1-5 i x p^2}{\sqrt{1-4 i x p^2}} ) \Big) ~~, 
\end{array}} \nonumber
\end{align}
while the moment $\nu_3 \equiv \Big\langle (qa)^3 \Big\rangle$ is:
\begin{align}
\nu_{3} = & \, \left( 1+3p^2+3 p^4 \right) x^3 ~. \nonumber
\end{align}

Then, the relations (\ref{f11})-(\ref{f12}) allows the calculation of the expressions of $f_{11}, f_{12}, f_{22}$, and the static structure factor $S(q)$ of the inverse Gaussian distribution population of spherical particles is obtainable from (\ref{Sfin0}).

\end{itemize}

\section{Comparing S(q) for the three previous cases to their analytical PY solutions}

\begin{figure}
\vspace*{-\baselineskip}
\centering
\includegraphics[scale=0.9]{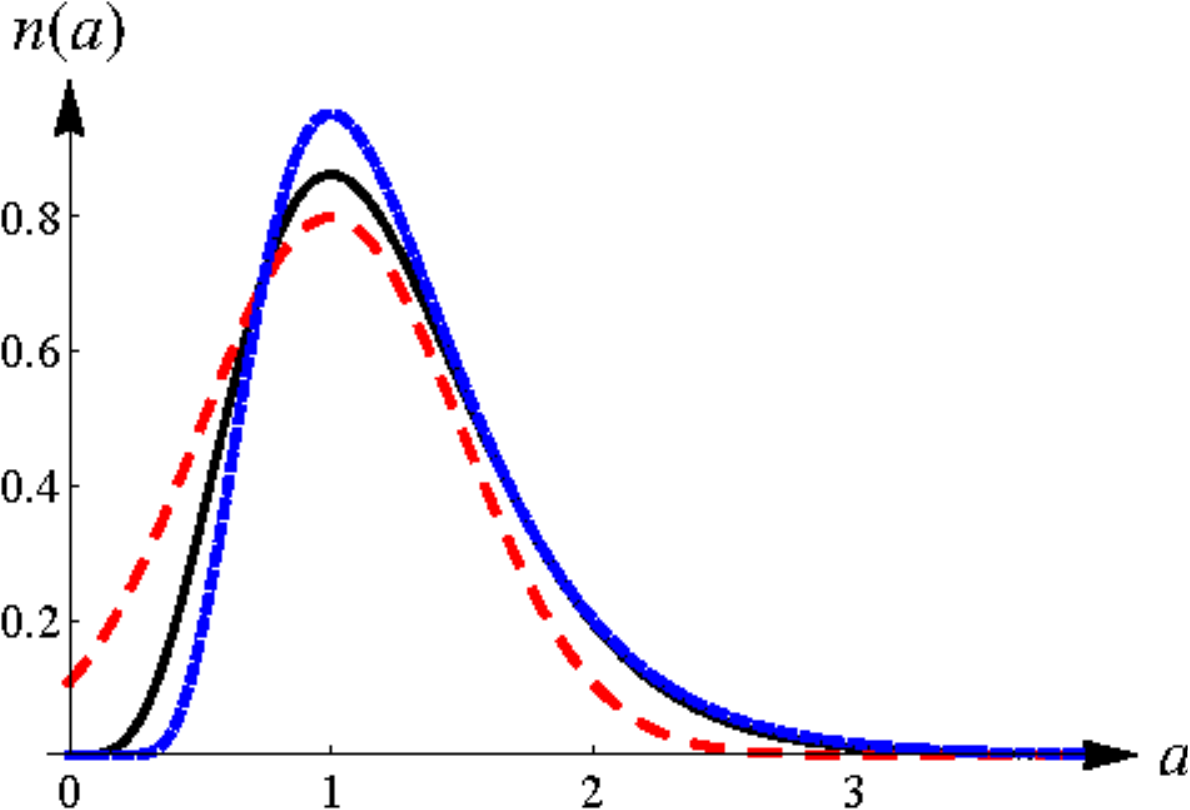}
\includegraphics[scale=0.8]{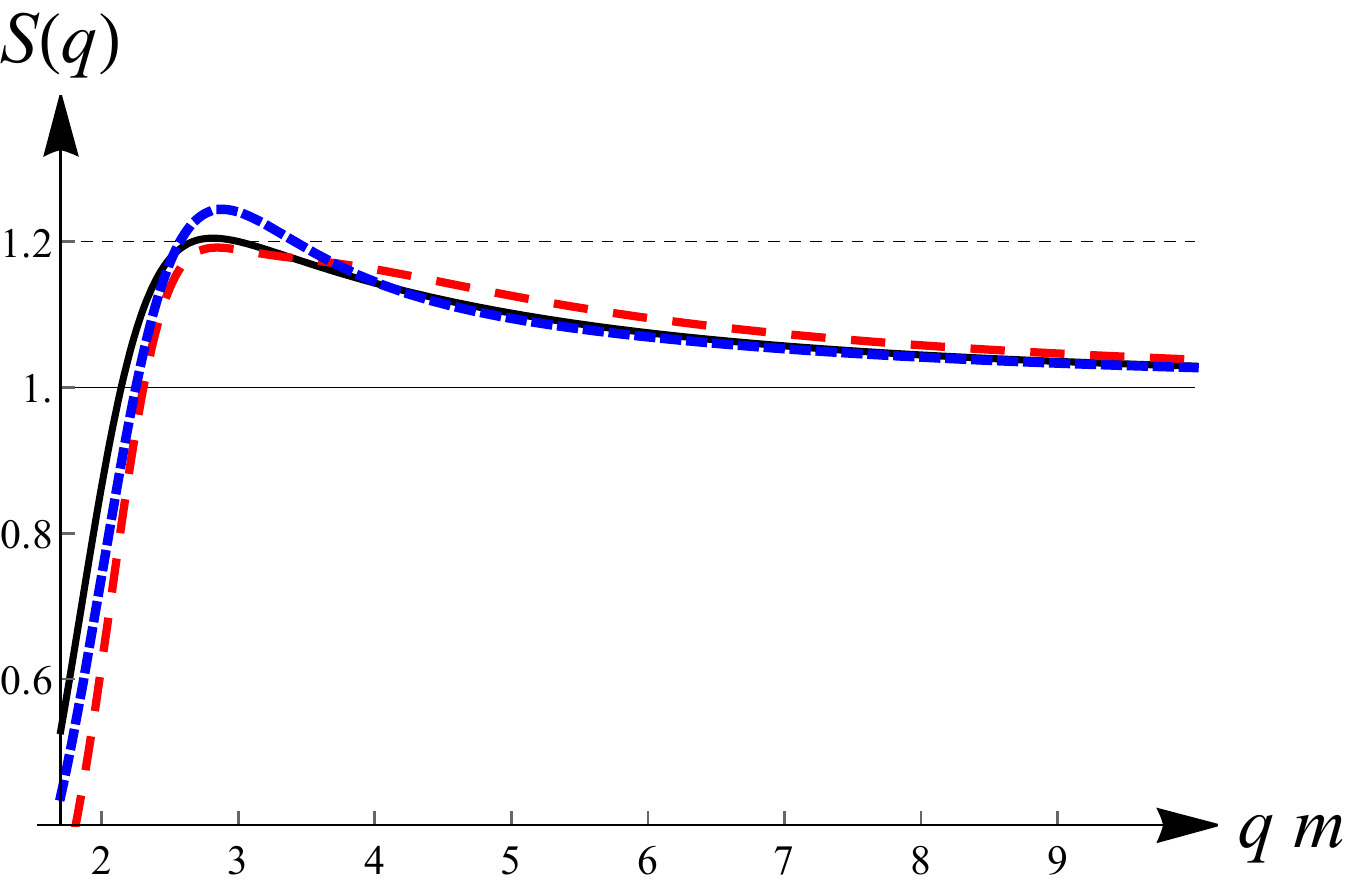}
\caption{(top) Three kinds of radius-distribution with same mode ($= 1$) and same standard deviation ($= 1/2$). The black continuous curve is Schulz distribution, the red dashed curve is truncated normal, and the blue dotted curve is inverse Gaussian distribution; (bottom)  respective structure factor functions in the scaled variable $q m$ (with $m = 1$ the common mode), for the three radius distributions shown above (same colour conventions as top figure). The horizontal lines $1$ and $1.2$ are drawn as guides.   
}
\label{fig2}
\end{figure}

FIG. \ref{fig2} presents an example of the static structure factors $S(q)$ at the volume fraction $\phi = 0.5$ for the three distributions above. The distributions are all of the same mode, centered on $1$, with the same standard deviation, $\sigma = 1/2$.

Although the three $S(q)$ behave in a very similar way because of their common parameters, one may nonetheless notice minute differences: 1)  Schulz and truncated normal distributions give  about the same $S_{\text{max}}$ ($\simeq 1.20$ in this case), while the peak is slightly higher ($\simeq 1.24$) for the inverse Gaussian distribution as a result of its sharper small-$x$ cut-off ; 2) the shape of the structure factor after the main peak decreases more slowly for the truncated normal distribution (Gaussian large-$x$ tail) than for the two other distributions (exponential large-$x$ tails).

\begin{figure}
\vspace*{-\baselineskip}
\centering
\includegraphics[scale=0.9]{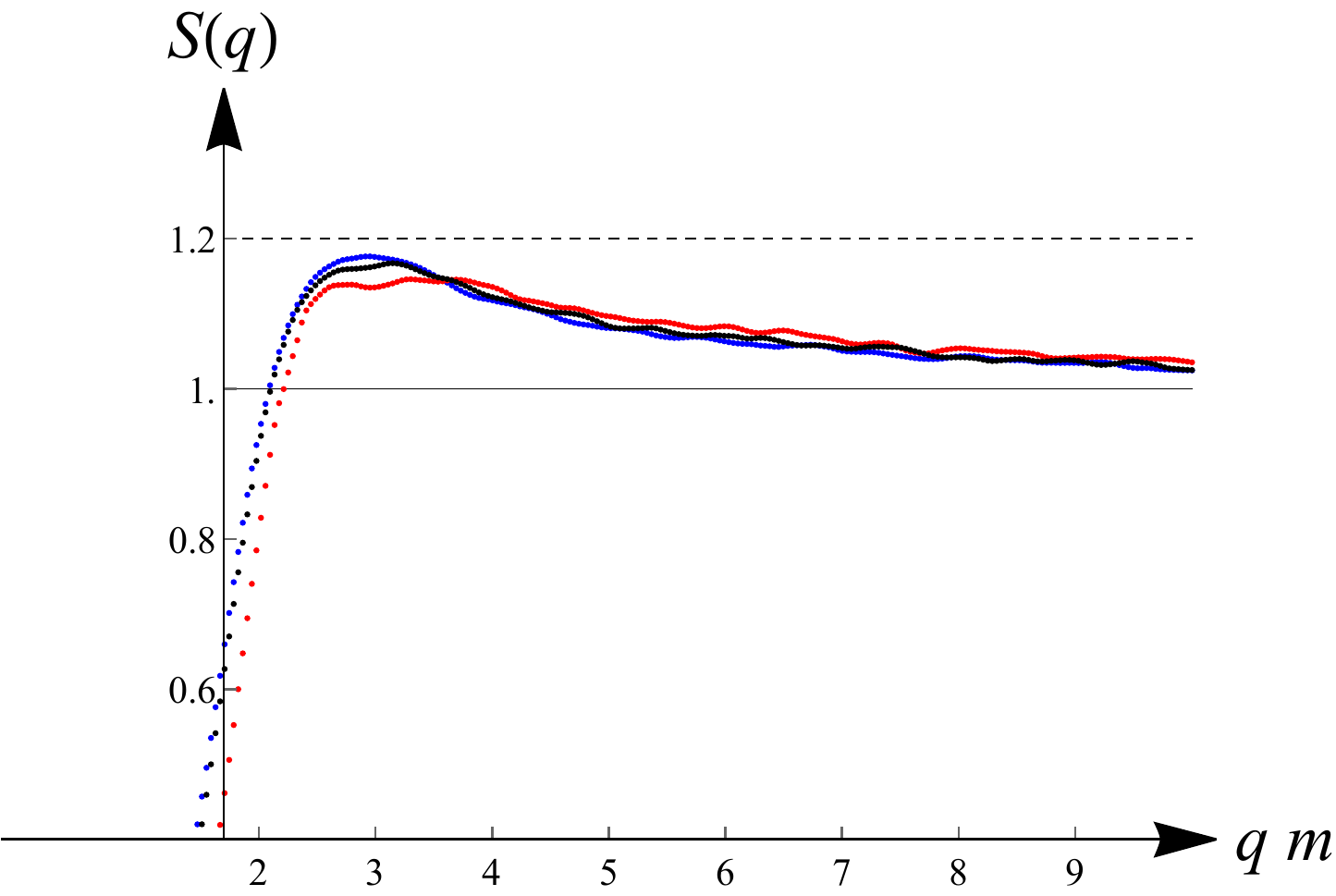}
\caption{Structure factor functions in the scaled variable $q m$ (with $m = 1$ the mode), for numerical systems made of $N = 10^6$ non-overlapping spheres with radii distributed according to Schulz, truncated normal and inverse Gaussian distributions (same colour conventions as  Fig.\ref{fig2}), at $\phi = 0.5$. Each curve is averaged over two independent samples. These curves can be compared with the analytical results shown on Fig\ref{fig2} with the same scales.  
}
\label{fig3}
\end{figure}

Fig.\ref{fig3} shows structure factors of numerical systems made of $N = 10^6$ particles at $\phi = 0.50$ for the three distributions reported in Fig.\ref{fig2}. Their shapes are very similar to the analytical  PY solutions, except that the PY-approximation leads to an overestimation by a few percent of the numerical $S_{\text{max}}$. The overestimation is similarly present and well-documented for monodisperse hard-sphere systems at high concentrations (Frenkel et al, 1986; Hansen \& McDonald, 2006).

\section{Exact $S(q)$ in the PY approximation for an Apollonian packing of spheres}

A very different kind of radius distribution appears in the problem of filling  space  totally with non-overlapping spherical particles. Such a packing requires using a population of spheres with power-law radius distribution (Stumpf \& Mason, 2012): 
\begin{align}
n(a) = 1/a^{d_f+1}  ~,  \label{na}
\end{align}
with the exponent: $2 < d_f < 3$ (Kinzel \& Reents, 1998). The most popular example of the distribution (\ref{na})  is the Apollonian packing (Mandelbrot, 1983) (for which $d_f \simeq 2.47$ (Borkovec et al, 1994), an iterative process in which the largest possible sphere is placed into the largest void left in the packing. Other  space-filling sphere packings are known with larger values of $d_f$ (Lieb \& Lebowitz, 1972). The  distribution (\ref{na}) is also found in other contexts, for example, the size distribution of natural aerosols, and the distribution is then often called the Junge distribution (Junge, 1955). 

In all practical applications -- experimental or numerical -- the values of the particle radius are confined between two limits (Varrato \& Foffi, 2011),  $a_{\text{min}}$ and $a_{\text{max}}$. Mathematically, these two limits ensure that the moments of orders $0, \cdots, 3$ of $n(a)$ exist and are finite. Experimentally, these two limits depend on the conditions in which the system is constructed. The non-dimensional parameter $\rho \equiv a_{\text{min}}/a_{\text{max}}$ plays a central role in the theory. In particular, the smaller the value of $\rho$, the denser the sphere packing. This relation can be formalized via the asymptotic ($\rho \ll 1$) expression of the porosity (Varrato \& Foffi, 2011):
\begin{align}
1 - \phi = C \, \rho^{3-d_f} ~, \label{Aste}
\end{align}
in which $\phi$ is the actual volume fraction of the packing, and $d_f$ the exponent of the power law (\ref{na}). The positive coefficient $C$ may depend on the value of $d_f$. The value $C \simeq 0.85$ holds for a  random Apollonian packing.

Distribution (\ref{na}) is written here under an unnormalized form, and we shall use this form in the following demonstrations. Indeed, as explained in the Section  \ref{Intro}, the average quantities appearing in the definitions of the auxiliary functions $b, c, d, e, f, g$, do not require normalization of the radius distribution.

The expressions (\ref{b})-(\ref{g}) to calculate the auxiliary quantities $b, c, d, e, f, g$ all depend on the quantity $3\phi/((1-\phi)\Big\langle (q a)^3 \Big\rangle)$ which deserves some preliminary attention in the present case. Indeed, using the power-law distribution (\ref{na}), one obtains the relation:
\begin{align}
\Big\langle a^3 \Big\rangle =  \dfrac{d_f}{3-d_f} a_{\text{max}}^3 \,\rho^{d_f} \left( \dfrac{1- \rho^{3- d_f} }{1- \rho^{\,d_f} } \right) ~, \label{a3}
\end{align}
in which the quantity in parenthesis is very close to 1 when $\rho \ll 1$ since $d_f < 3$. Using (\ref{Aste}) and (\ref{a3}) one deduces that:
\begin{align}
\dfrac{3\phi}{1-\phi} \dfrac{(q a_{\text{min}})^3}{\Big\langle (q a)^3 \Big\rangle} \simeq \dfrac{3-d_f}{d_f}\dfrac{3\phi}{C} ~, 
\end{align}
which is a positive constant ($\simeq 0.75$ for the random Apollonian packing) independent of $q$. \\

For the general power-law distribution, we use below the reduced variable:
\begin{align}
x \equiv 2q a_{\text{min}} \simeq \frac{2(d_f-1)}{d_f} \, q \Big\langle a \Big\rangle ~, 
\end{align}
and consider the situation where $\rho \ll 1$.

The functions $b, c, d, e, f, g$ express in terms of the two auxiliary functions $\cal I$ and $\cal J$ defined as:
\begin{align}
{\cal I}(x) \equiv & \, \text{si}(2-d_f,x) ~, \label{I} \\
{\cal J}(x) \equiv & \,  \dfrac{1}{(d_f-2)x^{d_f-2}}-\text{ci}(2-d_f,x)  ~,  \label{J}
\end{align}
in which  the two generalized sine and cosine integrals $\text{si}$ and $\text{ci}$ are (Olver et al, 2010):
\begin{align}
\text{si}(\lambda,x) \equiv  \int_{x}^{\infty} t^{\lambda -1}\sin t dt & ~~;~~\text{ci}(\lambda,x) \equiv \int_{x}^{\infty} t^{\lambda -1} \cos t dt ~, \nonumber 
\end{align}
defined for any $\lambda < 1$, and $x >0$. 
We find:
\begin{align}
b = & \dfrac{\alpha \phi}{d_f} \left( 4\dfrac{\sin x - x \cos x}{x^{d_f}}+(4-d_f) {\cal I}(x) \right) ~, \label{bpl} \\
c = & \dfrac{\alpha \phi}{d_f} \left( 4\dfrac{1+x^2/2- \cos x-x \sin x}{x^{d_f}}+(4-d_f) {\cal J}(x) \right) ~, \label{cpl} \\
d = & \, 1+\alpha  \phi \,{\cal I}(x) ~, \label{dpl} \\
e = & \, \alpha  \phi \, {\cal J}(x) ~, \label{epl} \\
f = & \,  \dfrac{\alpha \phi}{d_f-1} \left(2 \dfrac{1-\cos x}{x^{d_f-1}}+(3-d_f) {\cal I}(x) \right) ~, \label{fpl} \\
g = & \, \dfrac{\alpha \phi}{d_f-1} \left( 2\dfrac{x-\sin x}{x^{d_f-1}}+(3-d_f) {\cal J}(x) \right) ~, \label{gpl}
\end{align}
with 
\begin{align}
\alpha = 3 \, \dfrac{3-d_f}{C} \label{alpha} ~, 
\end{align}
the value of which is typically in the range of $1 - 3$. 

As usual, the structure factor function is obtained using the formulae (\ref{S1})-(\ref{Y}). \\

One can remark that the PY approximation leads here to a structure factor which has a definite non-trivial limit function for the densest system  $\phi = 1$ (that is one can safely replace $\phi$ by the value $1$ in (\ref{bpl})-(\ref{gpl})). This is because  even if space is entirely filled with material spheres, X-Ray scattering takes place at the fractal interface between the spheres. 

The Complementary Percus-Yevick approach discussed in the Section \ref{CPY} cannot apply in this case because the interface matter/void is strongly correlated (it is a fractal surface of fractal dimension $d_f$).

\subsection*{Comparison with the experimental structure factor of an extremely dense emulsion}

We will compare this analytical form to structure factors obtained from numerical experiments and from recent empirical experiments on extremely dense emulsions made of polydisperse spherical droplets.



\begin{figure}
\vspace*{-\baselineskip}
\centering
\includegraphics[scale=0.45]{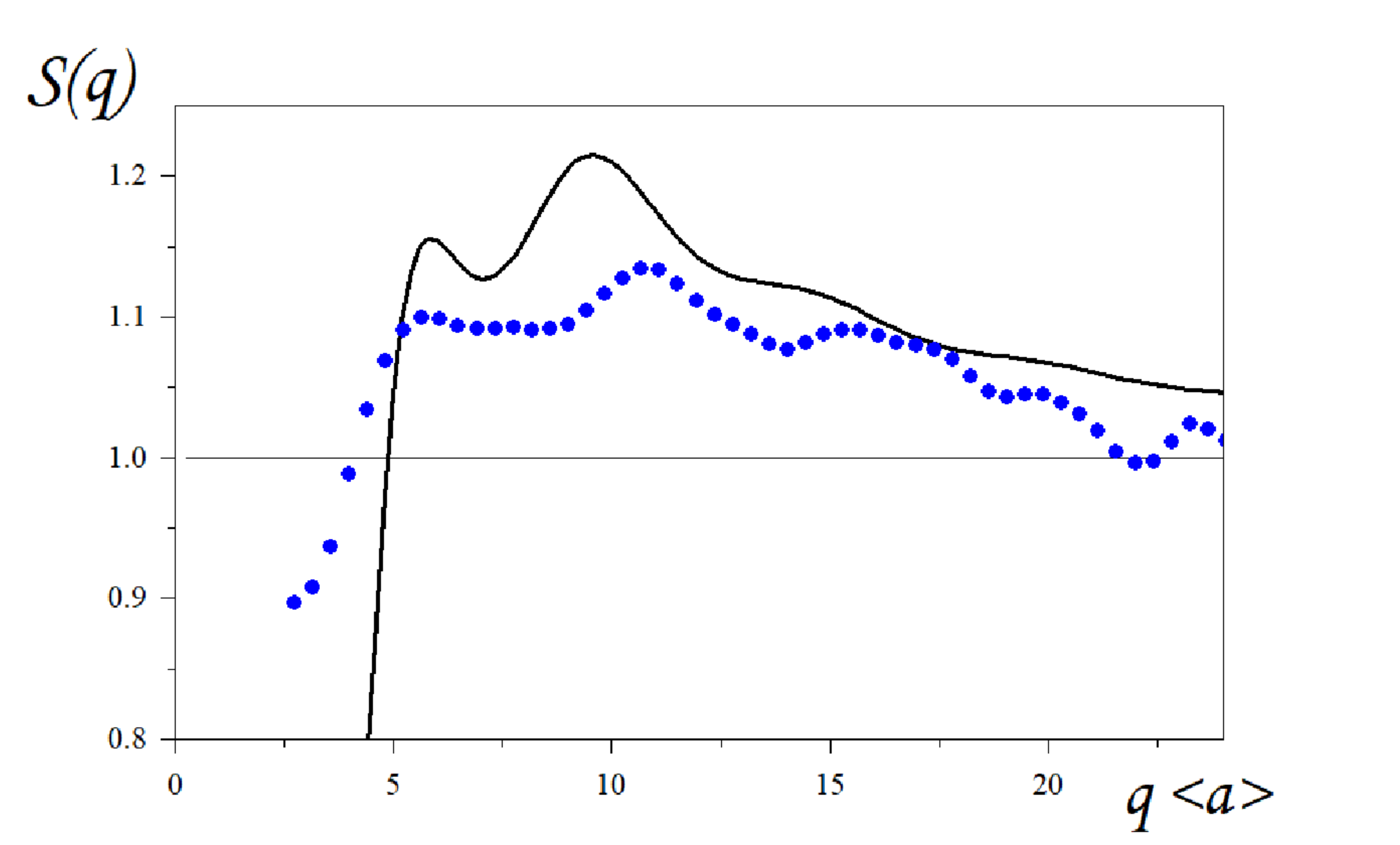}
\caption{(blue dots) Structure factor (blue dots) obtained by Small-Angle X-ray Scattering of a High Internal-Phase-ratio Emulsion made of spherical oil droplets packed at $\phi = 0.95 $ in water containing a limited amount of surfactant C12E6 (full details in the reference (Kwok et al, 2020)). The experimental $S(q)$ is compared to the  solution obtained from (\ref{bpl})-(\ref{gpl}) and formulae (\ref{S1})-(\ref{Y}), with the Apollonian packing parameters: $d_f = 2.47$,  $\alpha = 1.9$ and $\phi = 1$ (black continuous curve). The shape and  peak magnitudes are quite similar in both cases, though the PY solution leads to  peaks higher than the experimental ones by $5 \%$. }
\label{fig1a}
\end{figure}

Application of our approach is shown in FIG. \ref{fig1a} where the blue dots are experimental values of $S(q)$ obtained from a high internal-phase ratio emulsion  ($\phi = 0.95$) of oil in water in the presence of a small amount of C12E6 surfactant. The initial stirring speed was 250 rpm and the system was analysed after one month's evolution at rest. The full details of the experimental conditions and results of the Small-Angle X-ray Scattering data are given in (Kwok et al, 2020). 
For comparison, the black curve in FIG. \ref{fig1a} is the asymptotic solution (\ref{bpl})-(\ref{gpl}) for $\phi \rightarrow 1$ in the Apollonian packing case, that is:  $d_f = 2.47$ and $\alpha = 1.9$ ($\alpha$ is the parameter defined in (\ref{alpha})). The experimental value of $\Big\langle a \Big\rangle$ was the only parameter to have been adjusted, namely: $\Big\langle a \Big\rangle = 1.07 \mu$m, quite close to the value deduced from the experimental radius distribution, namely: $1.30 \mu$m. The overall shapes of both $S(q)$ functions are clearly similar with at least two peaks at about the same locations, namely $q \Big\langle a \Big\rangle \simeq 5$, and $q \Big\langle a \Big\rangle \simeq 10-11$. 

As usual with the PY approximation, one can notice systematic overestimation of the analytical values of $S(q)$ compared with the experimental data.

\section{Extension to polydisperse Yukawa particles}

We address now the question of dispersions of charged spheres interacting under a Yukawa potential. The range of the  interaction is measured by the Debye length, $1/\kappa$. 

As a standard approximation, the spatial structure of the system behaves as if each radius $a$ was replaced by its effective particle radius: $a + 1/\kappa$. The parameter $\psi$, as defined in (\ref{psi}) by: $\psi = 3\phi/(1-\phi)$, must then be modified by changing the actual value of the volume fraction $\phi$ to $\phi \Big\langle (a+1/\kappa)^3 \Big\rangle/\Big\langle a^3 \Big\rangle$. It results in the following new definition of $\psi$:
\begin{align}
\psi =  \, \dfrac{3\phi}{\phi_x-\phi} \label{newphi}
\end{align}
instead of (\ref{psi}) with the parameter $\phi_x = \Big\langle (\kappa a)^3 \Big\rangle/ \Big\langle (\kappa a+1)^3 \Big\rangle$ that depends on the radius distribution and not on the volume fraction. Thus, it follows for:
\begin{itemize}
\item Schulz distribution (\ref{Sc}) of parameters $s$ and $\Big\langle a \Big\rangle$:
\begin{align}
\phi_x =  \, \dfrac{1}{1+\dfrac{(1+3 \kappa \Big\langle a \Big\rangle)s^2+3(s+1)(\kappa \Big\langle a \Big\rangle)^2}{(s+1)(s+2)(\kappa \Big\langle a \Big\rangle)^3}}  \label{ds} 
\end{align}
\item truncated normal distribution (\ref{no}) of parameters $p$ and $\Big\langle a \Big\rangle$:
\begin{align}
\phi_x =  \, \dfrac{\kappa \Big\langle a \Big\rangle}{1+\kappa \Big\langle a \Big\rangle}\dfrac{1}{1+\dfrac{1+2\kappa \Big\langle a \Big\rangle}{(1+3p^2)(\kappa \Big\langle a \Big\rangle)^2}}  \label{dg} 
\end{align}
\item inversed Gaussian distribution (\ref{invg}) of parameters $p$ and $\Big\langle a \Big\rangle$:
\begin{align}
\phi_x =  \, \dfrac{1}{1+\dfrac{1+3\kappa \Big\langle a \Big\rangle+3(1+p^2)(\kappa \Big\langle a \Big\rangle)^2}{(1+3p^2+3p^4)(\kappa \Big\langle a \Big\rangle)^3}}  \label{di} 
\end{align}
\item power-law distribution (\ref{na}):
\begin{align}
\phi_x \simeq \dfrac{1}{1+\dfrac{3-d_f}{(\kappa a_{\text{max}})^3} \left( \dfrac{a_{\text{max}}}{a_{\text{min}}} \right)^{d_f}} \label{dp}
\end{align} 
\end{itemize}
 
Note that the value of $\phi_x$ for any radius distribution simplifies to $\phi_x = 1$ when the Debye length vanishes ($\kappa \rightarrow \infty$), that is, the uncharged particle case. 

As a result of the form of (\ref{newphi}), $\phi_x$ represents the larger volume fraction of the system for which the present approach can be used. Subsequently, the power-law radius distribution may be problematic when large volume fractions are involved. Indeed, $a_{\text{min}}$ tends to 0 as the volume fraction goes to 1. The expression (\ref{dp}) shows that in this case, $\phi_x$ may become much smaller than 1 when 
$a_{\text{min}}/a_{\text{max}} < 1/(\kappa a_{\text{max}})^{3/d_f}$. 
This condition must then be checked before applying the PY approximation for such a population of charged spheres.
\\

We give now an example of possible application of these formulae. Let us consider  the data and results published in (Cabane et al, 2016) on aqueous dispersions of spherical silica particles with a broad monomodal radius distribution:
\begin{align}
\text{polydispersity:}~~ p = & \, 14\% ~ , \label{pexp} \\
\text{mean radius:} ~\Big\langle a \Big\rangle = & \, 8 ~\text{nm}~ , \label{aexp} \\
\text{Debye length:} ~1/\kappa = & \, 4.5 ~\text{nm} ~. \label{kexp}
\end{align}

With these experimental parameters, the value of $\phi_x \simeq 0.271$ in the three cases above (\ref{ds})-(\ref{di}). This means in particular that our approach can be used in the limited range of $0 < \phi < 0.271$ (consistent with the experimental investigation range $0.038 \le \phi \le 0.24$). 

\begin{figure}
\includegraphics[scale=0.8,trim=80mm 40mm 40mm 40mm,clip]{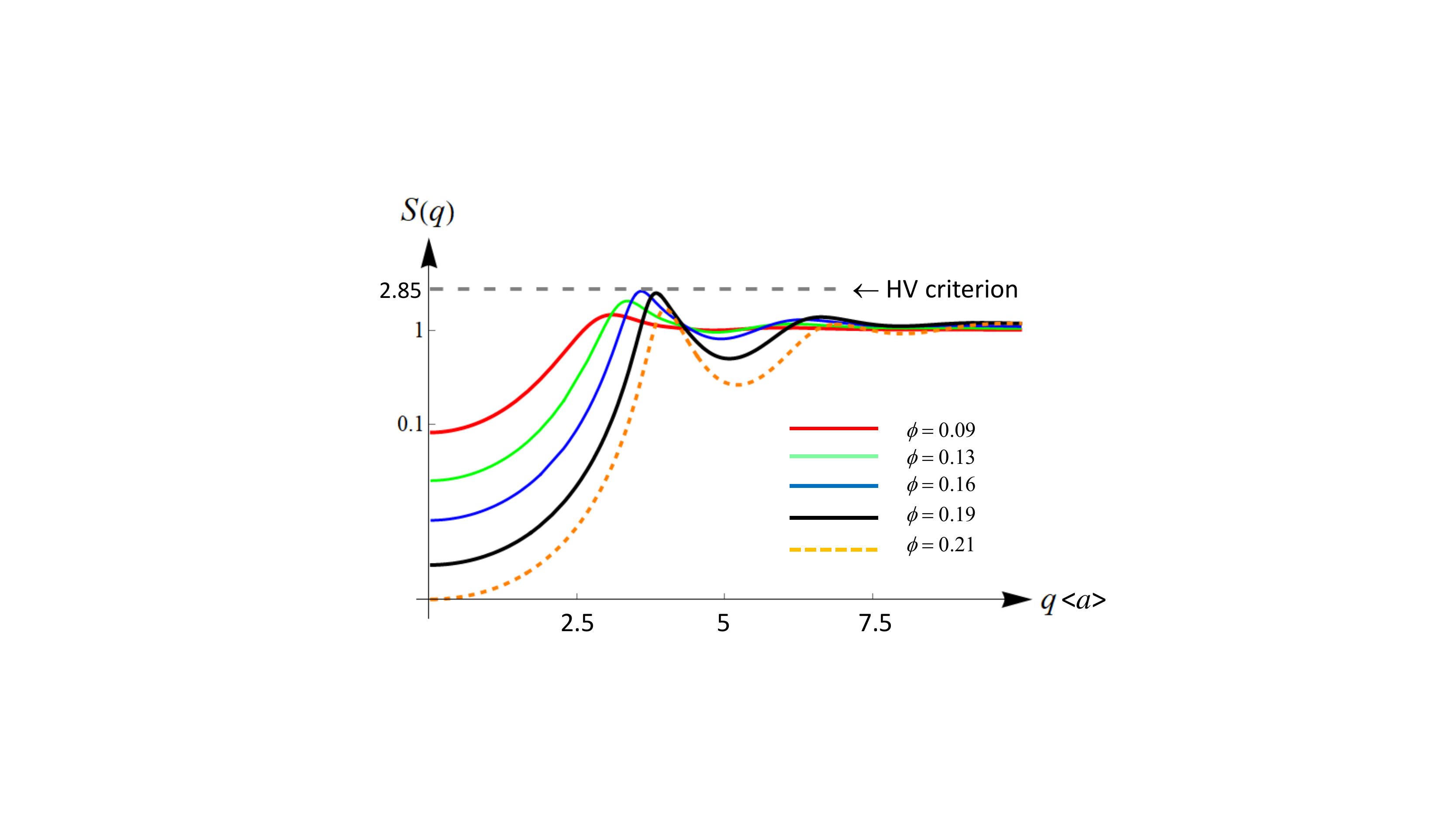}
\caption{Percus-Yevick structure factors for the truncated normal radius-distribution of charged spheres with parameters (\ref{pexp})-(\ref{kexp}) and five values of the volume fraction, $\phi = 0.09, 0.13, 0.16, 0.19, 0.21$. The height of the first peak increases regularly with $\phi$ up to the value of the Hansen-Verlet (HV) threshold 2.85. This is the expected sign for onset of crystallization. For the present parameters, it occurs  near $\phi = 0.16$ (blue continuous line). Beyond this value, the peak height reduction  demonstrates the decrease of the spatial short-range order in the particle positions. All these results compare perfectly with the experimental data shown in (Cabane et al, 2016). In this example, the Vrij's solution predicts quantitatively the volume fraction threshold where crystallization starts to occur in the system.   } 
\label{figLuc}
\end{figure}

On Fig. \ref{figLuc}, the structure factor $S(q)$ is plotted for the truncated normal distribution (\ref{no}), and five values of the volume fraction, $\phi = 0.09, 0.13, 0.16, 0.19, 0.21$.
The height $S_{\text{max}}$ of the first peak reflects the spatial order of the  particles in the system. As the volume fraction is increased,  the short-range order becomes stronger, and $S_{\text{max}}$ becomes higher. This occurs until the value of $S_{\text{max}}$  rises up to the threshold $2.85$: this is the Hansen-Verlet criterion for the onset of crystallization (Baus, 1983). Beyond this threshold, the rapid decrease of short-range order reveals jamming or clustering (limited aggregation) of the particles.
Examining the behaviour of the structure factor as calculated from our PY approach, one can expect crystallization to occur at $\phi \simeq 0.173$ which agrees perfectly with the experimental results (see Fig. 1a of (Cabane et al, 2016)). 

The case study here explored clearly demonstrates the great aptitude of of Vrij's PY solution to predict the liquid-solid transition in a system made of (possibly charged) polydisperse spheres.

\section{Conclusion}

Experimental structure factors measured by Small-Angle Scattering is a very powerful tool for obtaining information about the spatial distribution of colloidal particles in a dispersion. However, the interpretation of acquired spectra requires that one has definite models available for comparison. In this work, we have outlined a systematic and generalizable protocol for calculating the structure factor of any population of hard spheres at a given volume fraction and a given radius distribution (defined over any number of bins). In particular, we have derived and provided the pertinent analytical equations for several commonly encountered distributions in the domain of colloidal sciences (e.g. exponential, Schulz, normal, inverted Gaussian and power law). We further demonstrated our Complementary Percus-Yevick approach applicable to situations in which existing models fall short at very high volume fractions made possible by extreme polydispersities. It is thus our hope that our work may aide experimentalists in their Small-Angle Scattering data analysis through the ad-hoc creation of specifically useful comparison models that may not readily be available in literature.

\appendix

\section{Reformulation of the Vrij solution for the structure factor of an assembly of hard spheres}

We consider a two-phase system made of non-deformable spheres (e.g. phase O) dispersed in a liquid (phase W). The overall volume fraction of  phase O is $\phi$. Each sphere is supposed  homogeneous and spherical, and its radius is noted $a$. The population of polydisperse spherical particles has a  normalized radius-distribution $n(a)$ (normalization condition: $\int_0^{\infty} n(a) da = 1$).

Let $q$ be the  magnitude of the scattering vector in a small-angle scattering experiment. 
The normalized intraparticle interference factor of a homogeneous sphere of radius $a$, is noted: $\Phi(q a)$, with the function $\Phi(x) = 3(\sin x - x \cos x)/x^3$. The scattering amplitude contributed by the spherical particle of radius $a$ is then: $\alpha_3 a^3 \Phi(q a)$, in which $\alpha_3$ is a coefficient dependent on the refractive indices of the two phases.
 We also have to introduce the function $\Psi(x) =\sin x/x$. \\
 
The static structure factor, $S(q)$, of such a system is generally written as:
\begin{align}
S(q) = \frac{\phi_0}{\phi} \frac{I_{\phi}(q)}{I_{\phi_0}(q)}~,  \label{RR0}
\end{align}
in which $I_{\phi}(q)$ is the normalized intensity scattered by the system at the volume fraction $\phi$, and $I_{\phi_0}(q)$ the scattering intensity of the same system when diluted to an extremely small volume fraction $\phi_0 \simeq 0$. In (\ref{RR0}), the coefficient $\phi/\phi_0$ is the dilution factor. 

In the following steps, we use notations close to (Vrij, 1979), except for the definition of the averaged values of quantities related to the particle radii. Below, the averaged value of a quantity $f(a)$ is the standard one, namely:
\begin{align}
\Big\langle f(a) \Big\rangle \equiv \int_0^{\infty} f(a) \, n(a) da ~. \label{def<>}
\end{align}

The analytical results obtained by Vrij for  a polydisperse population of spheres are: 
\begin{align}
R_{\phi}(q) = & \frac{-D_f(q)}{\Delta(q)} ~,  \\
-\frac{(1-\phi)^4}{\phi} D_f(q) = & A_{0}  \big( \Big\langle a^6 \Phi^2 \Big\rangle |T_1 +  T_2|^2+ \Big\langle a^4 \Psi^2 \Big\rangle |T_3|^2 + \nonumber \\
+ &  \Big\langle a^5 \Phi \Psi \Big\rangle ((T_1+T_2) T_3^\star+(T_1^\star+T_2^\star) T_3)  \big) ~, \\
(1-\phi)^4 \Delta(q) = &  |T_1|^2 ~, 
\end{align}
in which $A_{0}$ is a normalization coefficient independent of $q$ and of $\phi$ but dependent on the electromagnetic properties of the emulsion and of the radius distribution. To be precise,  $A_0 = 16 \alpha_3^2/(4\pi \Big\langle a^3 \Big\rangle/3)$. The auxiliary functions $T_1, T_2, T_3$ are given by the following expressions (Vrij, 1979):
\begin{align}
F_{11} = & 1- \phi \left(1- \frac{\Big\langle a^3 \Phi e^{iqa} \Big\rangle}{\Big\langle a^3 \Big\rangle} \right) ~, \label{F11} \\
F_{22} = & 1- \phi \left(1- 3 \frac{\Big\langle a^3 \Psi e^{iqa} \Big\rangle}{\Big\langle a^3 \Big\rangle} \right) ~, \label{F22} \\
F_{12} = & \phi \frac{\Big\langle a^4 \Phi e^{iqa} \Big\rangle}{\Big\langle a^3 \Big\rangle} ~, \label{F12} \\
F_{21} = & (1 - \phi) iq - 3\phi  \frac{\Big\langle a^2 \Big\rangle - \Big\langle a^2 \Psi e^{iqa} \Big\rangle}{\Big\langle a^3 \Big\rangle} ~, \label{F21} \\
T_1 = & F_{11}F_{22}-F_{12}F_{21} ~, \label{T1} \\
T_2 = & \phi \left( F_{21} \frac{\Big\langle a^4 \Phi e^{iqa} \Big\rangle}{\Big\langle a^3 \Big\rangle} - F_{22} \frac{\Big\langle a^3 \Phi e^{iqa} \Big\rangle}{\Big\langle a^3 \Big\rangle} \right) ~, \label{T2} \\
T_3 = & 3\phi \left( F_{12} \frac{ \Big\langle a^3 \Phi e^{iqa} \Big\rangle}{\Big\langle a^3 \Big\rangle} - F_{11} \frac{\Big\langle a^4 \Phi e^{iqa} \Big\rangle}{\Big\langle a^3 \Big\rangle} \right) ~. \label{T3} 
\end{align}
 
These expressions lead  to the structure factor function under the form:
\begin{align}
S(q) = & \frac{(1-\phi)^2}{ |T_1|^2} \left(  |F_{22}|^2+ 9 \frac{\Big\langle a^4 \Psi^2 \Big\rangle}{\Big\langle a^6 \Phi^2 \Big\rangle} |F_{12}|^2 - \right. & \nonumber \\
-& \left. 3 \frac{\Big\langle a^5 \Phi \Psi \Big\rangle}{\Big\langle a^6 \Phi^2 \Big\rangle} (F_{22} F_{12}^\star+F_{22}^\star F_{12}) \right) ~, 
\end{align}
where we used the relations derived from  (\ref{F11}), (\ref{F12}), (\ref{T1}), (\ref{T2}):
\begin{align}
T_1 + T_2 = & ~~~~~ (1- \phi) F_{22} \label{T12} ~, \\
T_3 = & - 3(1-\phi) F_{12} ~. \label{T33}
\end{align}

To continue the calculation, it is now convenient to introduce the following functions:
\begin{align}
f_{11} = & \, 1+\psi \dfrac{s_0 + s_1}{\nu_3} ~, \label{f11A} \\
f_{22} = & \, 1+\psi \dfrac{s_2}{\nu_3} ~, \label{f22A} \\
f_{12} = & \, \psi \dfrac{s_1}{\nu_3} ~, \label{f12A}
\end{align}
where:
\begin{align}
\psi = & \frac{3 \phi}{1-\phi } ~, \label{psiA} \\
\nu_3 = & \Big\langle (q a)^3 \Big\rangle ~, \label{nu30} \\
\Xi(x) = & \sin x - x \cos x \label{Phi} ~, \\
s_0 = & \Big\langle \Xi (qa)  \, e^{i q a} \Big\rangle ~, \label{s0A} \\
s_1 = & -i\Big\langle  q a\,  \Xi (qa) \,  e^{i q a} \Big\rangle ~, \label{s1A} \\
s_2 = & \Big\langle (q a)^2 \, \sin (qa) \,  e^{i q a} \Big\rangle ~. \label{s2A} 
\end{align}
That way, one obtains the identity:
\begin{align}
|T_1|^2 = (1-\phi)^4 \left| f_{11} f_{22}+f_{12}^2 \right|^2 ~, 
\end{align}
hence the expression of the structure factor:
\begin{align}
S(q) = \frac{\Big\langle \Xi^2 \Big\rangle |f_{22}|^2+ \text{Im}\{s_2 \} |f_{12}|^2-i\text{Re}\{s_1 \} (f_{22}^\star f_{12} - f_{22} f_{12}^\star)}{\Big\langle \Xi^2 \Big\rangle \left| f_{11} f_{22}+f_{12}^2 \right|^2} ~. 
\end{align}
At last, one uses the following identities:
\begin{align}
\psi \Big\langle \Xi^2 \Big\rangle = & \text{Im}\{f_{11} \} ~, \label{p1} \\
\psi \text{Im}\{s_2 \} = & \text{Im}\{f_{22} \} ~, \label{p2} \\
\psi \text{Re}\{s_1 \} = & \text{Re}\{f_{12} \} ~, \label{p3} 
\end{align}
to put the expression of the structure factor under the compact form:
\begin{align}
S(q) = \frac{\text{Im}\{f_{22}^\star (f_{11} f_{22} + f_{12}^{\,2})\}}{\text{Im}\{f_{11} \} |f_{11} f_{22} + f_{12}^{\,2}|^{\,2}}~,  \label{Sfin0}
\end{align}
in which the three auxiliary complex-valued functions $f_{11}, f_{22}, f_{12}$ are defined by:
\begin{align}
f_{11} = & \, 1+ \psi~ \frac{\mu_1}{\nu_3} ~, \label{f11} \\
f_{22} = & \, 1 + \psi~ \frac{\mu_2}{\nu_3} ~, \label{f22} \\
f_{12} = & \, \psi~ \frac{\mu_3}{\nu_3} ~, \label{f12}
\end{align}
with the parameter $\psi$ given in (\ref{psiA}).
Generally, the parameters $\mu_1, \mu_2, \mu_3, \nu_3$ do not depend on the volume fraction $\phi$ of the scattering matter. They are given explicitly by the following formulae:
\begin{align}
\mu_{1} = & \, \Big\langle (1-i q a) \, (\sin(q a)- q a \cos (q a))\, e^{i q a} \Big\rangle ~, \label{mu1} \\
\mu_{2} = & \,  \Big\langle (q a)^2 \,\sin ( q a) \, e^{i q a}  \Big\rangle ~, \label{mu2} \\
\mu_{3} = & \, - i \Big\langle  qa \, (\sin(q a)- q a \cos (q a)) \,e^{i q a}  \Big\rangle ~, \label{mu3} \\
\nu_{3} = & \, \Big\langle (q a)^3 \Big\rangle ~. \label{nu3} 
\end{align}
An equivalent  form of the set of formulae (\ref{Sfin0})-(\ref{nu3}), using only real-valued functions,  is given in the Section \ref{Intro}. It results simply from rewriting (\ref{Sfin0}) with $f_{11} = \text{Re}\{f_{11}\}+ i \, \text{Im}\{f_{11}\}$ and the similar expressions for $f_{22}$ and $f_{12}$.

The auxiliary parameter $\mu_1$ has another role in the context of scattering intensity. Indeed, using the same quantities, the diluted scattering intensity of the polydisperse system, is:
\begin{align}
I_0 \propto \frac{\Big\langle \left(\sin (q a) - q a \cos (q a) \right)^2 \Big\rangle}{q^6} ~, 
\end{align}
in which the proportionality constant is independent of $q$ for a constant wavelength. 
Using (\ref{p1}), $I_0$ can be written under the form:
\begin{align}
I_0 \propto \frac{\text{Re} \{ \mu_{1} \} }{q^6 } ~. 
\end{align} 


\subsection*{comparison with published results}

\begin{figure}
	\includegraphics[width=\columnwidth] {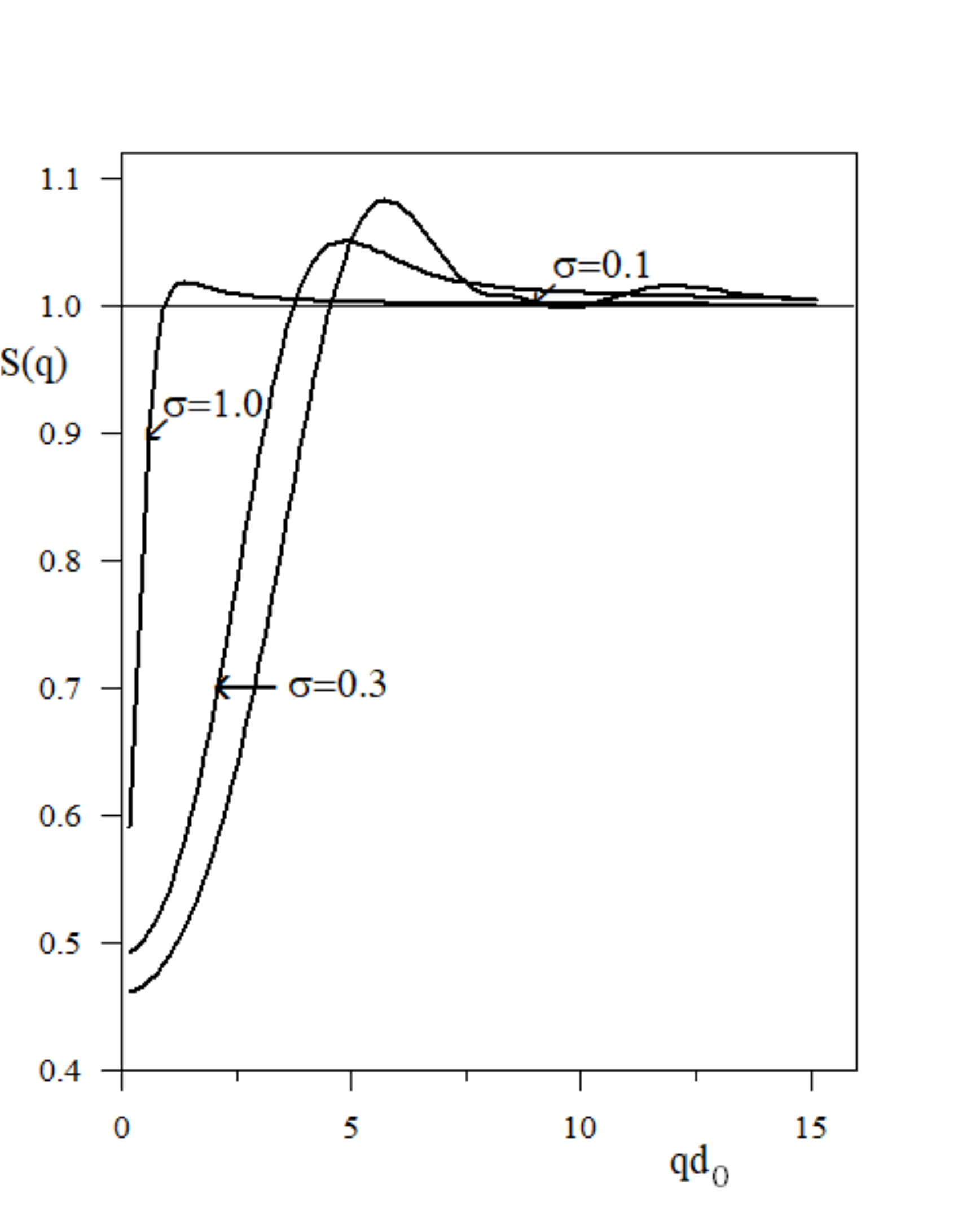}
    \caption{Structure factor $S(q)$ at volume fraction $\phi = 0.1$ for several standard deviations $\sigma$ of hard-sphere diameters. The notation $d_0 = 2 \Big\langle a \Big\rangle$ is used. These data are to be compared with  FIG. 4 of reference (van Beurten \& Vrij, 1981).}
    \label{fig4}
\end{figure}

\begin{figure}
	\includegraphics[width=\columnwidth] {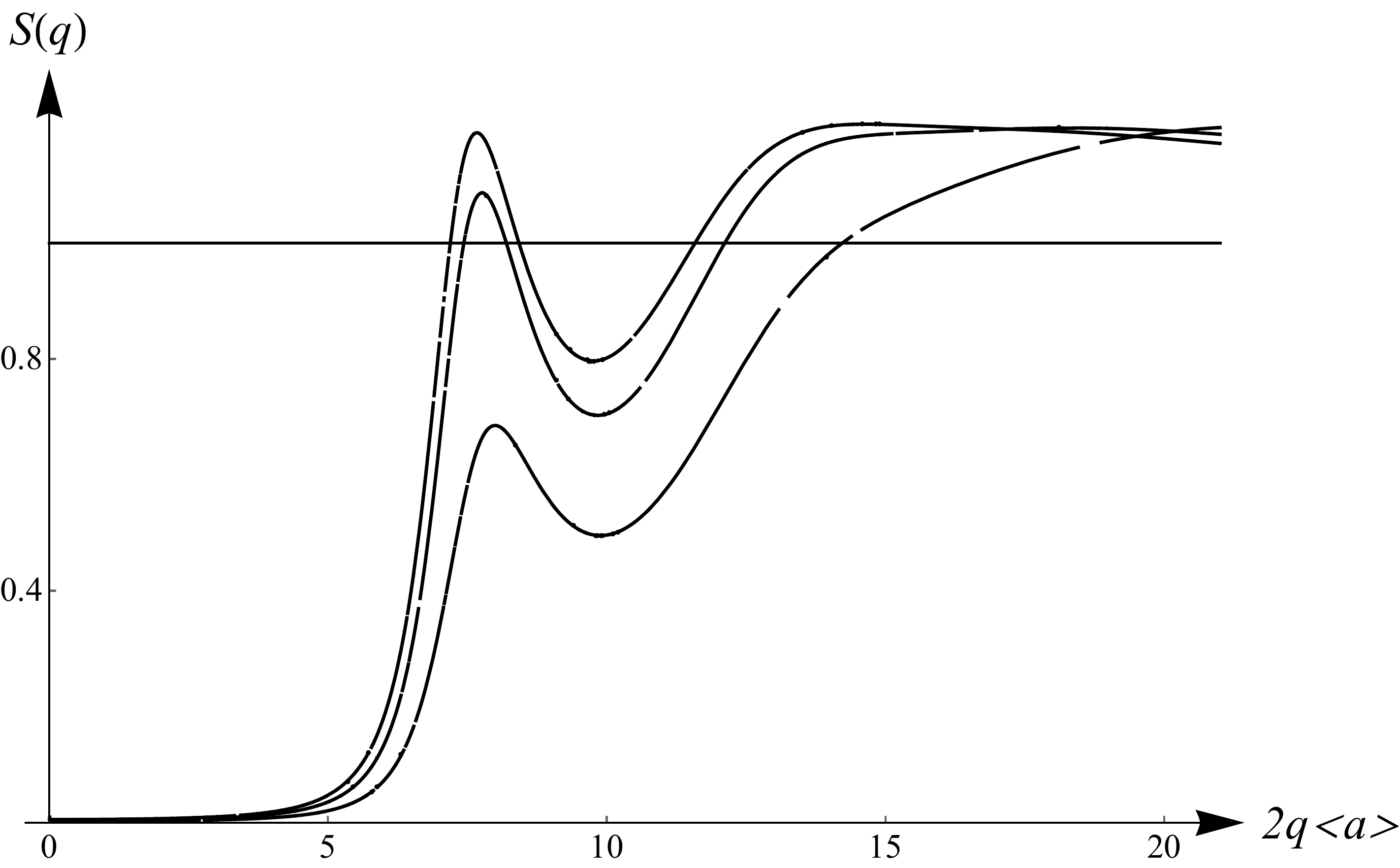}
    \caption{Scaled static structure factor function $S(q)$ at volume fractions and parameters: $(\phi = 0.7~;~s = 23.34)$, $(\phi = 0.72~;~ s=24.27)$, $(\phi  = 0.77~;~s = 22.68)$, respectively (from top to bottom). These data have to be compared with Figures 1.a), b), c) of reference (Scheffold \& Mason, 2009).}
    \label{figMason}
\end{figure}

\begin{itemize}

\item In (van Beurten \& Vrij, 1981), the authors showed structure factor functions calculated from an assembly of hard spheres with a Schulz distribution of the particle diameters and various values of  standard deviation. The original solution proposed by Vrij (Vrij, 1979) is used. Their FIG. 4, for example, corresponds to data for the volume fraction $\phi = 0.1$ with diameter standard deviations $= 0$ (monodisperse case), $0.1, 0.3$ and $1.0$. The Schulz distribution with mode $\Big\langle 2a \Big\rangle = 1$ is used, that is:
\begin{align}
n(a) \propto \left( a e^{-2a} \right)^{b-1} ~, 
\end{align}
with the following relation between the positive exponent $b$ and the standard deviation, $\sigma_a$, of the radius distribution:
\begin{align}
\sigma_a = \frac{\sqrt{b}}{2(b-1)} ~. 
\end{align} 
The monodisperse case corresponds to $b \rightarrow \infty$. 

The results obtained from the results given in the Section \ref{analy} for the Schulz distribution with  $\phi = 0.1$ and the three values of the standard deviations used in (van Beurten \& Vrij, 1981) are shown in our FIG. \ref{fig4}. The results are similar to   previously published data.
 
\item In (Scheffold \& Mason, 2009), the authors compared their experimental data to the analytical solution presented in (Ginoza \& Yasutomi, 1999) for the structure factor of a polydisperse system with a Schulz diameter distribution. Using the same set of parameters, one finds our FIG. \ref{figMason} which can be compared to their Figure 1.

\end{itemize}

\section{formulae for some  radius-distributions using the  complex-valued auxiliary quantities (\ref{mu1})-(\ref{mu3})}

\subsection{the Schulz distribution}

The case of  Schulz distribution: 
\begin{align}
\boxed{
n(a) \propto  ~ a^{s-1} e^{-s \, a/\langle a \rangle}} ~,  \label{Scp} 
\end{align}
is discussed in details in the Section \ref{analy}. We  give here the complete solution for $S(q)$ in terms of the complex-valued auxiliary functions $\mu_1, \mu_2, \mu_3$, in order to obtain a solution in a more compact form.

Let us introduce the auxiliary function $\mu_0$:
\begin{align}
\mu_0 = & \dfrac{e^{i(s+2) \tan^{-1} 2x}}{(1 + 4x^2)^{s/2+1}} ~, \nonumber
\end{align}
and the scaled variable $x = q \Big\langle a \Big\rangle/s$.

The auxiliary parameters $\mu_1, \mu_2, \mu_3$ are such that:
\begin{align}
\boxed{
\begin{array}{rcl}
\mu_{1} =  \, \dfrac{i}{2} \Big[ 1 + s (s+1) x^2  - ~~~~~~~~~~~~~~~~~~~~~~~~~~~~&  \\
-  \mu_0 \left(1 - 2(s+2) i x - (s+1) (s+4) x^2 \right) \Big] ~, \\[3mm] 
\mu_{2} =  \, \dfrac{i x^2}{2} s (s+1)   \left(1 - \mu_0 \right) ~,  ~~~~~~~~~~~~~~~~~~~~~~~~~~~\\[3mm] 
\mu_3 =  \, \dfrac{s x}{2} \Big(1+ i (s+1) x - \mu_0 (1 - i(s+3) x)  \Big) ~, ~~~
\end{array}} \nonumber
\end{align}
and
\begin{align}
\nu_{3} = & \,  s (s+1 ) (s+2 ) x^3 ~. \nonumber
\end{align}

One can notice that when the value of the parameter $s$ is an integer number, the expression of $\mu_0$ is quite simple:
\begin{align}
\mu_0 = \dfrac{1}{(1-2i x)^{s+2}} ~. \nonumber
\end{align}
Consequently, $S(q)$ can in this case be expressed as the ratio of two polynomials in $x$. For example:
\begin{align}
s = 1 & \Rightarrow \mu_0 = \dfrac{1-12x^2+2ix(3-4x^2)}{(1+4x^2)^3} ~, \label{s1} \\
s = 2 & \Rightarrow \mu_0 = \dfrac{1-24x^2+14x^4+8ix(1-4x^2)}{(1+4x^2)^4} ~, \label{s2} \\
\text{etc.} \nonumber
\end{align} 
and the static structure factor, $S(q)$,  is then obtained readily from expression (\ref{Sfin0}).

\subsection{the power-law distribution}

The power-law radius distribution is defined as:
\begin{align}
\boxed{
n(a) \propto \frac{1}{a^{d_f+1}} ~~~~\text{for}~~a_{\text{min}} < a < a_{\text{max}}}~. 
\label{power}
\end{align}
The exponent $d_f$ is here restricted to the values $2 < d_f < 3$.
The auxiliary positive parameter: $\rho \equiv a_{\text{min}}/a_{\text{max}}$ is used henceforth, introduced by the relation:
\begin{align}
1- \phi \simeq \kappa \, \rho^{3-d_f} ~, 
\end{align}
as conjectured in (Varrato \& Foffi, 2011), and $\kappa$ is a number with value of order 1.
At last, we define the parameter $w$ as:
\begin{align}
w = \dfrac{1-\phi}{\kappa -(1-\phi)} ~. \label{gw}
\end{align}

The values of the first moments of the radius are calculated:
\begin{align}
\Big\langle a \Big\rangle = &  \frac{d_f}{d_f-1}  \,  a_{\text{min}} \left( \dfrac{ 1 - \rho^{d_f-1}}{1 - \rho^{d_f} } \right) ~, \label{maxi3} \\
\Big\langle a^2 \Big\rangle =  & \frac{d_f}{d_f-2}  \,  a_{\text{min}}^2 \left( \dfrac{ 1 - \rho^{d_f-2}}{1 - \rho^{d_f}} \right) ~, \label{maxi4} \\
\Big\langle a^3 \Big\rangle = &  \frac{d_f}{3-d_f}  \,  a_{\text{min}}^3 \dfrac{1}{\rho^{3-d_f}} \left( \dfrac{ 1 - \rho^{3-d_f}}{1 - \rho^{d_f} } \right) ~. \label{maxi5} 
\end{align}
Note the term $\rho^{d_f-3}$ in expression (\ref{maxi5}), which may be quite large in the case of wide power-law distributions (that is when $\rho \ll 1$).

Defining the reduced variable:
\begin{align}
x \equiv q a_{\text{min}} ~, \label{pl0}
\end{align}
the expressions for $\mu_1, \mu_2, \mu_3$ can be expressed after introducing the auxiliary function $\mu_0$ written in terms of the incomplete Gamma function, namely:
\begin{align}
\mu_0(x) =  & \, \dfrac{\sin x}{x}e^{i x} + (-2 i x)^{d_f - 3} \Gamma( 3 - d_f, -2 i  x) \label{mu_0} ~, 
\end{align}

\begin{align}
\boxed{
\begin{array}{rcl}
\dfrac{\mu_k}{\nu_3} =   \, w \, \dfrac{3-d_f}{d_f-2} \, \Big[ \varphi_k(x) - \rho^{d_f-3} \varphi_k(x/\rho) \Big] ~~~~~;~~~~ k = 1, 2, 3 ~~~,  \\[3mm]
\varphi_1(x) =  \,i \, \dfrac{d_f-2}{2\, d_f} \left( \dfrac{1+2x^2-e^{2i x} (1 - 2 i x)}{x^3} \right) + ~~~~~~~~~~~~~~~~\nonumber \\[2mm]
+   \dfrac{4 - d_f}{d_f} \mu_0(x) ~, ~~~~~~~~~~~~~~~~~~~~~~~~~~~~~~~~~~~~~~~~~~~~~~~~~~  \\[4mm]
\varphi_2(x) =  \mu_0(x)~,  ~~~~~~~~~~~~~~~~~~~~~~~~~~~~~~~~~~~~~~~~~~~~~~~~~~~~~~~~~~~ \\[4mm]
\varphi_3(x) =  \, i \, \dfrac{d_f-2}{2(d_f-1)} \left( \dfrac{2x-i(1 - e^{2 i x})}{x^2} \right) +  \dfrac{3 - d_f}{d_f-1} \mu_0(x) ~.  ~~~~~
	\end{array}}
\end{align}
It is interesting to note that when $\rho \ll 1$, all the quantities $\rho^{d_f-3} \varphi_k(x/\rho)$ appearing in the above formulae are negligible, since $\varphi(x) \sim  1/x$ in $x \rightarrow 0$, for all the three indices $k = 1, 2, 3$, and $d_f >2$. In addition, all the quantities in parenthesis in the moments (\ref{maxi3})-(\ref{maxi5}) are $\simeq 1$. Taking into account these approximations, and taking the real values of all quantities, we can then recover the formulae (\ref{bpl})-(\ref{gpl}). 









\end{document}